\documentclass[12pt]{article}

\usepackage{amssymb}
\usepackage{graphicx}
\usepackage{amsthm}
\usepackage{psfrag}
\usepackage{dsfont}

\newtheoremstyle{custom}
{15pt}
{15pt}
{}
{}
{\bfseries}
{}
{.5em}
{}
\theoremstyle{custom}
\newtheorem{definition}{Definition}
\newtheorem{proposition}[definition]{Proposition}
\newtheorem{lemma}[definition]{Lemma}
\newtheorem{example}[definition]{Example}
\newtheorem{corollary}[definition]{Corollary}
\newtheorem{condition}[definition]{Condition}
  
\begin{document}

Published: Richard A Barry and Susan M Scott 2014 Class. Quantum Grav. 31(12) 125004 doi: 10.1088/0264-9381/31/12/125004
\\

\begin{center}\textbf{The Strongly Attached Point Topology of the Abstract Boundary For Space-Time}
\\

Richard A Barry and Susan M Scott
\\

\textit{Centre for Gravitational Physics, College of Physical and Mathematical Sciences, The Australian National University, Canberra ACT 0200, Australia}
\\

richard.barry@anu.edu.au, Susan.Scott@anu.edu.au\end{center}

\begin{abstract} The abstract boundary construction of Scott and Szekeres provides a `boundary' for any $n$-dimensional, paracompact, connected, Hausdorff, $C^{\infty}$ manifold. Singularities may then be defined as objects within this boundary.  In a previous paper \cite{Barry}, a topology referred to as the attached point topology was defined for a manifold and its abstract boundary, thereby providing us with a description of how the abstract boundary is related to the underlying manifold.  In this paper, a second topology, referred to as the strongly attached point topology, is presented for the abstract boundary construction. Whereas the abstract boundary was effectively disconnected from the manifold in the attached point topology, it is very much connected in the strongly attached point topology. A number of other interesting properties of the strongly attached point topology are considered, each of which support the idea that it is a very natural and appropriate topology for a manifold and its abstract boundary.\end{abstract}

\\

AMS classification scheme numbers: 53C23, 57R40, 57R15, 83C75

\section{Introduction}

In the paper `The attached point topology of the abstract boundary for space-time' \cite{Barry}, a topology for a manifold $\mathcal{M}$ and its collection of abstract boundary points $\mathcal{B}(\mathcal{M})$ was constructed.  The topology, referred to as the attached point topology, represents one of the more natural topologies that can be placed upon the abstract boundary. It was produced via obvious extensions to the abstract boundary point definitions, and did not require any additional conditions to be placed upon the manifold or its boundary.  However, it was demonstrated that it was possible to separate the manifold and its abstract boundary by disjoint open sets of the attached point topology.  The manifold and its abstract boundary were therefore disconnected from one another in some sense.  Even so, the fact that the attached point topology was Hausdorff was a pleasing result and suggested that the attached point topology was a good starting point in producing a topology that is more descriptive of the topological relationship between a manifold and its abstract boundary.

Because the abstract boundary is produced via embeddings of the manifold, the abstract boundary exists in a space separate to that of the manifold.  A topology on $\mathcal{M}\cup\mathcal{B}(\mathcal{M})$ should therefore connect the abstract boundary to the underlying manifold $\mathcal{M}$.  As noted previously, the attached point topology provided one such description of the topological relationship between $\mathcal{M}$ and $\mathcal{B}(\mathcal{M})$.  This topology relied on the notion of abstract boundary points being `close', in some sense, to open sets of the manifold $\mathcal{M}$.  Abstract boundary points that were close to an open set of $\mathcal{M}$ were said to be `attached' to that open set, and thus the `location' of an abstract boundary point could be described relative to the known topology of $\mathcal{M}$.  Because it is possible to separate the manifold and its abstract boundary from one another by disjoint open sets of the attached point topology, it can be said that the attached point topology does not fully integrate $\mathcal{B}(\mathcal{M})$ with the structure of $\mathcal{M}$.  The fact that $\mathcal{M}$ and $\mathcal{B}(\mathcal{M})$ can be separated in this way is a result of there being too many open sets in the attached point topology.  In this paper, a new topology referred to as the strongly attached point topology is considered.  The strongly attached point topology is defined similarly to the attached point topology but has one additional restriction.  This restriction limits the `type' of open set in $\mathcal{M}$ to which an abstract boundary point may be considered to be `close to', and hence there are comparatively fewer open sets in the strongly attached point topology.  The main consequence of this is that every open neighbourhood of an abstract boundary point necessarily contains some part of the manifold $\mathcal{M}$, i.e., the abstract boundary is topologically inseparable from the underlying manifold $\mathcal{M}$.

In section \ref{abstract boundary}, the abstract boundary will again be defined as a matter of convenience. The strongly attached abstract boundary point definition is developed in \ref{Strongly attached Boundary Points and Sets}, which describes how an abstract boundary point may be related back to $\mathcal{M}$.  The strongly attached point topology, which utilises the previously mentioned definition is presented in section \ref{strongly attached topology section}. Various properties of the topology are then discussed in sections \ref{Open and Closed sets}, \ref{the inclusion map}, \ref{Contact properties} and \ref{optimal embeddings and partial cross sections}.

We refer the reader interested in the g-boundary, b-boundary and c-boundary to \cite{Geroch}, \cite{Schmidt} and \cite{Geroch1}. For those interested in the more recent causal boundary, see \cite{Senovilla}, \cite{c-boundary}, \cite{Flores}, \cite{flores1}, \cite{flores2} and \cite{flores3}.

Within this work, we use the following fact frequently and so formally present it here for ease of reference.  Let $g$ be a Riemannian metric on a manifold $\mathcal{M}$, and let $\Omega_{p,q}$ denote the set of piecewise smooth curves in $\mathcal{M}$ from $p$ to $q$.  For every curve $c\in\Omega_{p,q}$ with $c:[0,1]\rightarrow\mathcal{M}$ there is a finite partition $0=t_1<t_2<...<t_k=1$ such that $c\mid[t_i,t_{i+1}]$ is smooth for each $i$, $1\leq i\leq k-1$.  The Riemannian arc length of $c$ with respect to $g$ is then defined to be $L(c)=\sum_{i=1}^{k-1}\int_{t_i}^{t_{i+1}}\sqrt{g(c'(t),c'(t))}dt$, and the Riemannian distance function, $d(p,q)$, between $p$ and $q$ is then defined in terms of this by $d(p,q)=\textrm{inf}\{L(c):c\in\Omega_{p,q}\}\geq0$.  The most useful property of this distance function is that the open balls defined by $B_\epsilon(p)=\{q\in\mathcal{M}:d(p,q)<\epsilon\}$ form a basis for the manifold topology, and thus the topology induced by the Riemannian metric agrees with the manifold topology \cite{Lorentzian}.

\section{The Abstract Boundary}\label{abstract boundary}

For the convenience of the reader, we will provide the definition of the a-boundary in this section.  For a more complete discussion of the a-boundary, see \cite{Scott}, \cite{fama}, \cite{Ashley} and \cite{Whale}. It will be assumed that all manifolds used in the following work will be n-dimensional, paracompact, connected, Hausdorff and smooth (i.e., $C^{\infty}$).  The manifold topology will be employed throughout the paper unless explicitly stated otherwise.  The principle feature of the a-boundary construction is that of an envelopment.

\begin{definition}[Embedding]The function $\mathcal{\phi: M\rightarrow\widehat{M}}$ is an \textit{embedding} if $\phi$ is a homeomorphism between $\mathcal{M}$ and $\phi(\mathcal{M})$, where $\phi(\mathcal{M})$ has the subspace topology inherited from $\widehat{\mathcal{M}}$.\end{definition}

\begin{definition}[Envelopment]An \textit{enveloped manifold} is a triple ($\mathcal{M,\widehat{M},\phi}$) where $\mathcal{M}$ and $\mathcal{\widehat{M}}$ are differentiable manifolds of the same dimension $n$ and $\phi$ is a $\mathcal{C^{\infty}}$ embedding $\mathcal{\phi: M\rightarrow\widehat{M}}$.  The enveloped manifold will also be referred to as an \textit{envelopment of $\mathcal{M}$ by $\mathcal{\widehat{M}}$}, and $\mathcal{\widehat{M}}$ will be called the \textit{enveloping manifold.}\end{definition}

\begin{definition}[Boundary point]A \textit{boundary point p} of an envelopment ($\mathcal{M,\widehat{M},\phi}$) is a point in the topological boundary of $\phi(\mathcal{M})$ in $\mathcal{\widehat{M}}$.  The set of all such points p is thus given by $\partial(\phi(\mathcal{M}))=\overline{\phi(\mathcal{M})}\backslash\phi(\mathcal{M})$ where $\overline{\phi(\mathcal{M})}$ is the closure of $\phi(\mathcal{M})$ in $\mathcal{\widehat{M}}$.  The boundary points are then simply the limit points of the set $\phi(\mathcal{M})$ in $\widehat{\mathcal{M}}$ which do not lie in $\phi(\mathcal{M})$ itself.

The characteristic feature of a boundary point is that every open neighbourhood of it (in $\widehat{\mathcal{M}}$) has non-empty intersection with $\phi(\mathcal{M})$.\end{definition}

\begin{definition}[Boundary set]A \textit{boundary set B} is a non-empty set of such boundary points for a given envelopment, i.e., a non-empty subset of $\partial(\phi(\mathcal{M}))$.\end{definition}

It is important to note that different boundary points will arise with different envelopments of $\mathcal{M}$.  In order to continue, a notion of equivalence between boundary sets of different envelopments is required.  This equivalence is defined in terms of a covering relation.

\begin{figure}[htb!]
\centering%
\psfrag{manifold}{$\scriptstyle\widehat{\mathcal{M}}$}
\psfrag{e}{$\scriptstyle\phi(\mathcal{M})$}
\psfrag{f}{$\scriptstyle\phi'(\mathcal{M})$}
\psfrag{m}{$\scriptstyle\widehat{\mathcal{M}}$}
\psfrag{a}{$\scriptstyle\widehat{\mathcal{M}'}$}
\psfrag{g}{$\scriptstyle \phi \circ \phi'^{-1}(\mathcal{U'}\cap\phi'(\mathcal{M}))$}
\includegraphics{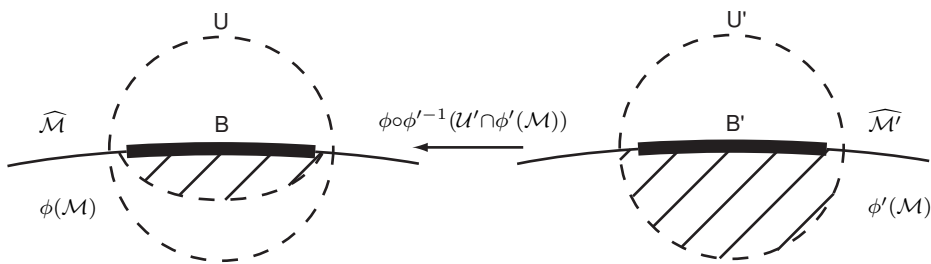}
\caption{the boundary set $B$ covers the boundary set $B'$}
\label{covering relation}
\end{figure}

\begin{definition}[Covering relation]Given a boundary set $B$ of one envelopment ($\mathcal{M,\widehat{M},\phi}$) and a boundary set $B'$ of a second envelopment ($\mathcal{M,\widehat{M'},\phi'}$), then \textit{$B$ covers $B'$}, denoted $B\triangleright B'$, if for every open neighbourhood $\mathcal{U}$ of $B$ in $\mathcal{\widehat{M}}$ there exists an open neighbourhood $\mathcal{U'}$ of $B'$ in $\mathcal{\widehat{M'}}$ such that \begin{displaymath} \phi \circ \phi'^{-1}(\mathcal{U'}\cap\phi'(\mathcal{M}))\subset\mathcal{U}. \end{displaymath}
In essence, this definition says that a sequence of points from within $\mathcal{M}$ cannot get close to points of $B'$ without at the same time getting close to points of $B$.  See figure \ref{covering relation}.\end{definition}

\begin{definition}[Equivalent]The boundary sets $B$ and $B'$ are \textit{equivalent} (written $B\sim B'$) if $B\triangleright B'$ and $B'\triangleright B$.  This definition produces an equivalence relation on the set of all boundary sets.  An equivalence class is denoted by $[B]$, where $B$ is a representative of the set of equivalent boundary sets under the covering relation.\end{definition}

\begin{definition}[Abstract boundary point and abstract boundary]An \textit{abstract boundary point} is defined to be an equivalence class $[B]$ that has a singleton boundary point $\{p\}$ as a representative member.  Such an equivalence class will then simply be denoted by $[p]$. The set of all such abstract boundary points of a manifold $\mathcal{M}$ will be denoted by $\mathcal{B(M)}$ and called the \textit{abstract boundary} of $\mathcal{M}$.  The union of all points of a manifold $\mathcal{M}$ and its collection of abstract boundary points $\mathcal{B(M)}$ may then be labelled as $\overline{\mathcal{M}}$, i.e., $\overline{\mathcal{M}}=\mathcal{M}\cup\mathcal{B}(\mathcal{M})$.\end{definition}

\begin{definition}[Covered abstract boundary point]An abstract boundary point $[p]$ \textit{covers} an abstract boundary point $[q]$, denoted $[p]\triangleright[q]$, if the representative singleton boundary point $\{p\}$ covers the representative singleton boundary point $\{q\}$. \end{definition}

\section{Strongly Attached Boundary Points and Sets}\label{Strongly attached Boundary Points and Sets}

The attached point topology was defined by topologically relating the abstract boundary points of a manifold $\mathcal{M}$ back to the points of $\mathcal{M}$ via the definition of an attached boundary point. We include this definition and the definition of an attached boundary set for the benefit of the reader.

\begin{definition}[Attached boundary point]\label{attached point}Given an open set $U$ of $\mathcal{M}$ and an envelopment $\phi:\mathcal{M}\rightarrow \widehat{\mathcal{M}}$, then a boundary point $p$ of $\partial (\phi(\mathcal{M}))$ is said to be \textit{attached to} $U$ if every open neighbourhood $N$ of $p$ in $\widehat{\mathcal{M}}$ has non-empty intersection with $\phi(U)$, i.e., $N\cap\phi(U)\neq\emptyset$.\end{definition}

\begin{definition}[Attached boundary set]\label{attached set} Given an open set $U$ of $\mathcal{M}$ and an envelopment $\phi:\mathcal{M}\rightarrow \widehat{\mathcal{M}}$, then a boundary set $B\subset\partial\phi(\mathcal{M})$ is said to be \textit{attached to} $U$ if every open neighbourhood $N$ of $B$ in $\widehat{\mathcal{M}}$ has non-empty intersection with $\phi(U)$, i.e., $N\cap\phi(U)\neq\emptyset$.\end{definition}

The strongly attached point topology also relies on the notion of an abstract boundary point being attached to an open set of $\mathcal{M}$, but the manner in which the abstract boundary point is attached is different.

\begin{figure}[htb!]
\centering%
\psfrag{manifold}{$\widehat{\mathcal{M}}$}
\psfrag{e}{$\phi(\mathcal{M})$}
\psfrag{b}{$N\cap\phi(\mathcal{M})$}
\psfrag{a}{$\phi(U)$}
\psfrag{n}{$N$}
\includegraphics{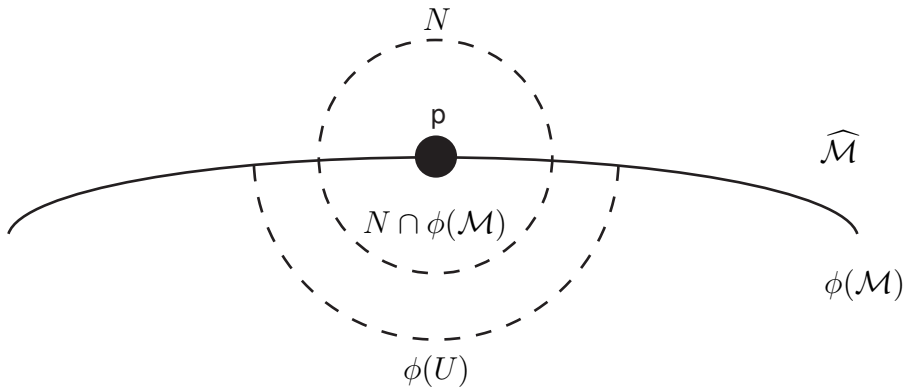}
\caption{a boundary point $p$ strongly attached to the open set $U$}
\label{strongly attached}
\end{figure}

\begin{definition}[Strongly attached boundary point]\label{strongly attached point} Given an open set $U$ of $\mathcal{M}$ and an envelopment $\phi:\mathcal{M}\rightarrow \widehat{\mathcal{M}}$, then a boundary point $p$ of $\partial(\phi(\mathcal{M}))$ is said to be \textit{strongly attached to} $U$ if there exists an open neighbourhood $N$ of $p$ in $\widehat{\mathcal{M}}$ such that $N\cap\phi(\mathcal{M})\subseteq\phi(U)$.  See figure \ref{strongly attached}.\end{definition}

\begin{lemma}\label{strong=attached point}If the boundary point $p$ is strongly attached to the open set $U$ then $p$ is attached to $U$.\end{lemma}

\textit{Proof:} Consider an envelopment $\phi:\mathcal{M}\rightarrow\widehat{\mathcal{M}}$ and a boundary point $p\in\partial(\phi(\mathcal{M}))$.  Suppose that $p$ is strongly attached to the open set $U\subset\mathcal{M}$.  There therefore exists an open neighbourhood $N$ of $p$ in $\widehat{\mathcal{M}}$ such that $N\cap\phi(\mathcal{M})\subseteq\phi(U)$.  Any other open neighbourhood of $p$ will have non-empty intersection with $N\cap\phi(\mathcal{M})$.  This follows from the fact that the intersection of two open sets is another open set: $N'$ is an open set that contains $p$, and thus $N\cap N'=N^*$ is an open set that also contains $p$.  Because $N^*$ is a neighbourhood of the boundary point $p$ we have that $N^*\cap\phi(\mathcal{M})\neq\emptyset$.  This implies that $(N\cap N')\cap\phi(\mathcal{M})\neq\emptyset$, and so $N'\cap\phi(U)\neq\emptyset$.  This is a statement of the attached boundary point condition, i.e., $p$ is attached to $U$.  $\Box$
\\

The requirement that there exists an open neighbourhood $N$ of $p$ in $\widehat{\mathcal{M}}$ such that $N\cap\phi(\mathcal{M})\subseteq\phi(U)$ removes the possibility of boundary points being strongly attached to open sets like those depicted in figure \ref{attached point}, i.e., open sets that are shaped more like a wedge and which have minimal `contact' with the boundary of the particular envelopment. If a boundary point is strongly attached to an open set $U$ then that set $U$ will always be more `spread out' along the boundary under the given envelopment.

\begin{figure}[htb!]
\centering%
\psfrag{m}{$\widehat{\mathcal{M}}$}
\psfrag{e}{$\phi(\mathcal{M})$}
\psfrag{a}{$\phi(U)$}
\psfrag{N}{$N$}
\includegraphics{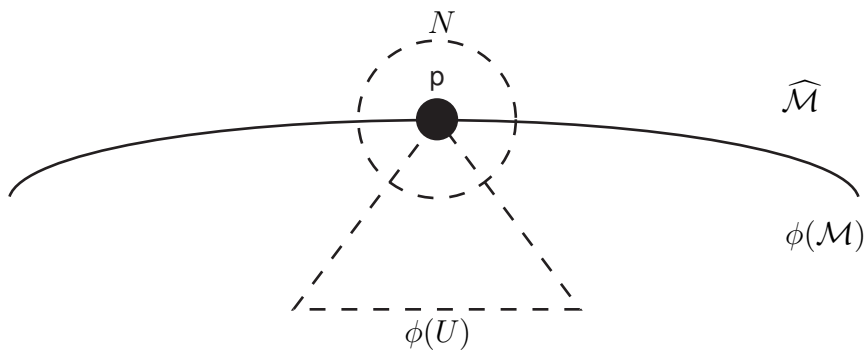}
\caption{a boundary point $p$ attached, but not strongly attached, to an open set $U$ of $\mathcal{M}$}
\label{attached point}
\end{figure}

\begin{figure}[htb!]
\centering%
\psfrag{manifold}{$\widehat{\mathcal{M}}$}
\psfrag{e}{$\phi(\mathcal{M})$}
\psfrag{n}{$N\cap N'$}
\psfrag{a}{$\phi(U_1)$}
\psfrag{b}{$\phi(U_2)$}
\includegraphics{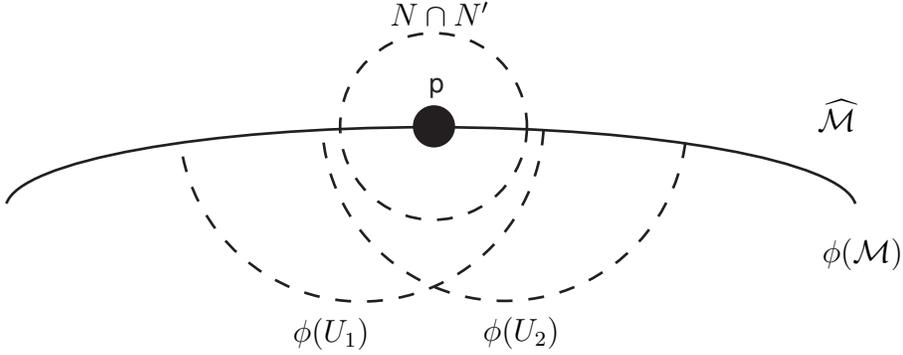}
\caption{the boundary point $p$ is strongly attached to $U_1\cap U_2$}
\label{stongly attached to u1 and u2}
\end{figure}

\begin{lemma}\label{strong intersection}If a boundary point $p\in\partial(\phi(\mathcal{M}))$ is strongly attached to the open sets $U_1$ and $U_2$, then $U_1\cap U_2\neq\emptyset$.  Furthermore, $p$ is strongly attached to $U_1\cap U_2$.  See figure \ref{stongly attached to u1 and u2}. \end{lemma}

\textit{Proof:} The boundary point $p$ is strongly attached to $U_1$ and so there exists an open neighbourhood $N$ of $p$ in $\widehat{\mathcal{M}}$ such that $N\cap\phi(\mathcal{M})\subseteq\phi(U_1)$.  It is also strongly attached to $U_2$ and so there exists an open neighbourhood $N'$ of $p$ in $\widehat{\mathcal{M}}$ such that $N'\cap\phi(\mathcal{M})\subseteq\phi(U_2)$. Now $N$ and $N'$ are both open neighbourhoods of $p$ and so their intersection is another open neighbourhood of $p$.  In addition, $p$ is a boundary point, and so every open neighbourhood of $p$ has non-empty intersection with $\phi(\mathcal{M})$.  We therefore have $(N\cap N')\cap\phi(\mathcal{M})\neq\emptyset$.  Now, since $N\cap\phi(\mathcal{M})\subseteq\phi(U_1)$ and $N'\cap\phi(\mathcal{M})\subseteq\phi(U_2)$, we have that $(N\cap N')\cap\phi(\mathcal{M})\subseteq\phi(U_1)$ and $(N\cap N')\cap\phi(\mathcal{M})\subseteq\phi(U_2)$.  This implies that $\phi(U_1)\cap\phi(U_2)\neq\emptyset$ and therefore that $U_1\cap U_2\neq\emptyset$.  Moreover, $(N\cap N')\cap\phi(\mathcal{M})\subseteq\phi(U_1)\cap\phi(U_2)=\phi(U_1\cap U_2)$ from which it follows that $p$ is strongly attached to $U_1\cap U_2$. $\Box$

\begin{figure}[htb!]
\centering%
\psfrag{manifold}{$\widehat{\mathcal{M}}$}
\psfrag{e}{$\phi(\mathcal{M})$}
\psfrag{n}{$N$}
\psfrag{u}{$\phi(U)$}
\psfrag{o}{$N\cap\phi(\mathcal{M})$}
\includegraphics{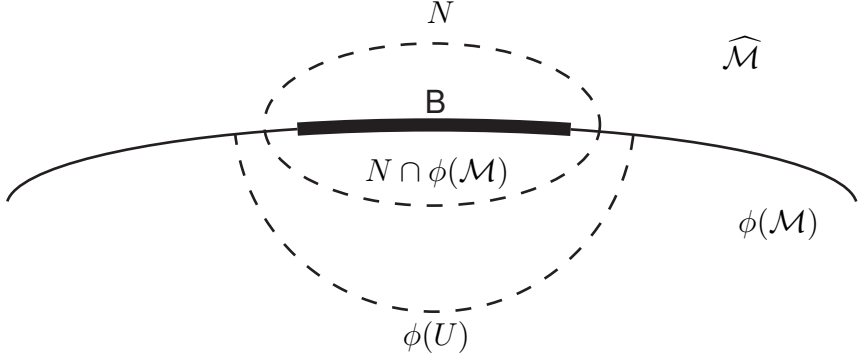}
\caption{a boundary set $B$ is strongly attached to an open set $U$ of $\mathcal{M}$}
\label{strongly attached boundary}
\end{figure}

\begin{definition}[Strongly attached boundary set]\label{strong=attached set}Given an open set $U$ of $\mathcal{M}$ and an envelopment $\phi:\mathcal{M}\rightarrow \widehat{\mathcal{M}}$, then a boundary set $B\subset\partial(\phi(\mathcal{M}))$ is said to be \textit{strongly attached to} $U$ if there exists an open neighbourhood $N$ of $B$ in $\widehat{\mathcal{M}}$ such that $N\cap\phi(\mathcal{M})\subseteq\phi(U)$.  See figure \ref{strongly attached boundary}.\end{definition}

\begin{lemma}\label{strong implies attached}If $B\subset\partial(\phi(\mathcal{M}))$ is strongly attached to the open set $U\subset\mathcal{M}$ then $B$ is attached to $U$.\end{lemma}

\textit{Proof:} The proof of this is identical to the proof of lemma \ref{strong=attached point}, except that we are dealing with open neighbourhoods of a boundary set rather than open neighbourhoods of a boundary point.  $\Box$

\begin{lemma}\label{all q attached}A boundary set $B\subset\partial(\phi(\mathcal{M}))$ is strongly attached to an open set $U\subseteq\mathcal{M}$ if and only if every boundary point $p\in B$ is strongly attached to $U$.\end{lemma}

\textit{Proof:} ($\Rightarrow$) Let $\phi:\mathcal{M}\rightarrow\widehat{\mathcal{M}}$ be an envelopment, and $B\subset\partial(\phi(\mathcal{M}))$ a boundary set that is strongly attached to the open set $U\subseteq\mathcal{M}$.  Because $B$ is strongly attached to $U$, there exists an open neighbourhood, $N$, of $B$ such that $N\cap\phi(\mathcal{M})\subseteq\phi(U)$.  $N$ is an open neighbourhood of every boundary point $p\in B$.  Clearly then, every $p\in B$ is strongly attached to $U$.
\\

($\Leftarrow$) If every boundary point $p\in B$ is strongly attached to $U$, then there exists an open neighbourhood, $N_p$, of each $p$ such that $N_p\cap\phi(\mathcal{M})\subseteq\phi(U)$.  The union $N_B$ of every $N_p$, i.e., $N_B=\bigcup_{p\in B} N_p$, is an open neighbourhood of $B$ such that $N_B\cap\phi(\mathcal{M})\subseteq\phi(U)$.  The boundary set $B$ is therefore strongly attached to $U$.  $\Box$

\begin{lemma}\label{strong intersecting set}If a boundary set $B$ is strongly attached to the open sets $U_1$ and $U_2$, then $U_1\cap U_2\neq\emptyset$. Furthermore, $B$ is strongly attached to $U_1\cap U_2$.\end{lemma}

\textit{Proof:} The proof follows from lemma \ref{strong intersection}, except that we are dealing with open neighbourhoods of a boundary set, rather than open neighbourhoods of a boundary point.  $\Box$
\\

Because boundary points which are equivalent may appear in a number of different envelopments, it is necessary to check that definitions \ref{strongly attached point} and \ref{strong=attached set} are well defined under the equivalence relation.  More specifically, we wish to show that if a boundary set $B\subset\partial(\phi(\mathcal{M}))$ is strongly attached to an open set $U\subset\mathcal{M}$ and there exists a boundary set $B'\subset\partial(\psi(\mathcal{M}))$ that is equivalent to $B$, then $B'$ is also strongly attached to $U$.

\begin{proposition}\label{B strongly attached}Let $B\subset\partial(\phi(\mathcal{M}))$ be strongly attached to an open set $U\subset\mathcal{M}$, and let $B'$ be a boundary set of a second envelopment $\phi':\mathcal{M}\rightarrow\widehat{\mathcal{M}'}$.  If $B\triangleright B'$, then $B'$ is also strongly attached to $U$.\end{proposition}

\textit{Proof:} The boundary set $B\subset\partial(\phi(\mathcal{M}))$ is strongly attached to the open set $U$.  There therefore exists an open neighbourhood $N$ of $B$ such that $N\cap\phi(\mathcal{M})\subseteq\phi(U)$.  Now, since $B\triangleright B'$, we have that $\phi\circ\phi'^{-1}(N'\cap\phi'(\mathcal{M}))\subset N$, where $N'$ is an open neighbourhood of $B'$ in $\widehat{\mathcal{M}'}$.  It follows that $\phi\circ\phi'^{-1}(N'\cap\phi'(\mathcal{M}))\subset N\cap\phi(\mathcal{M})\subseteq\phi(U)$, and therefore $N'\cap\phi'(\mathcal{M})\subseteq\phi'(U)$, i.e., $B'$ is strongly attached to $U$. $\Box$

\begin{definition}[Strongly attached abstract boundary point]\label{strongly attached abstract} The abstract boundary point $[p]$ is said to be \textit{strongly attached to} the open set $U$ of $\mathcal{M}$ if the boundary point $p$ is strongly attached to $U$.\end{definition}

The abstract boundary point $[p]$ is an equivalence class of boundary sets which are equivalent to $\{p\}$.  By proposition \ref{B strongly attached} the strongly attached abstract boundary point definition is well defined as any boundary set $B$ such that $B\sim p$ is also strongly attached to $U$, i.e., all members of the equivalence class $[p]$ are strongly attached to $U$.

Also, by lemma \ref{strong implies attached} it is clear that if an abstract boundary point $[p]$ is strongly attached to an open set $U$ of $\mathcal{M}$, then $[p]$ is also attached to $U$.

\begin{proposition}\label{BU closed}Consider an open set $U$ of $\mathcal{M}$ and an envelopment $\phi:\mathcal{M}\rightarrow\widehat{\mathcal{M}}$.  Let $B_U$ be the set of boundary points of $\partial(\phi(\mathcal{M}))$ which are strongly attached to $U$. The set $B_U$ is closed in $\widehat{\mathcal{M}}$ if and only if the limit points of $B_U$ are strongly attached to $U$.\end{proposition}

\textit{Proof:} Since $B_U\subset\partial(\phi(\mathcal{M}))$, any limit point of $B_U$ in $\widehat{\mathcal{M}}$ will also lie in $\partial(\phi(\mathcal{M}))$. By definition, $B_U$ is closed in $\widehat{\mathcal{M}}$ if and only if $B_U$ contains all its limit points. Clearly, $B_U$ contains all its limit points if and only if the limit points of $B_U$ are strongly attached to $U$, from which the result follows. $\Box$

\begin{corollary}Consider an open set $U$ of $\mathcal{M}$ and an envelopment $\phi:\mathcal{M}\rightarrow\widehat{\mathcal{M}}$.  Let $B_U$ be the set of boundary points of $\partial(\phi(\mathcal{M}))$ which are strongly attached to the $U$. The set $B_U$ is closed in the induced topology on $\partial(\phi(\mathcal{M}))$ if and only if the limit points of $B_U$ are strongly attached to $U$.\end{corollary}

\textit{Proof:} Since $\partial(\phi(\mathcal{M}))$ is closed in $\widehat{\mathcal{M}}$ and $B_U\subset\partial(\phi(\mathcal{M}))$, the set $B_U$ is closed in the induced topology on $\partial(\phi(\mathcal{M}))$ if and only if $B_U$ is closed in $\widehat{\mathcal{M}}$.  The result then follows directly from proposition \ref{BU closed}. $\Box$
\\

\begin{figure}[htb!]
\centering%
\psfrag{manifold}{$\widehat{\mathcal{M}}$}
\psfrag{e}{$\phi(\mathcal{M})$}
\psfrag{B}{$B_U$ (not closed)}
\psfrag{U}{$\phi(U)$}
\includegraphics{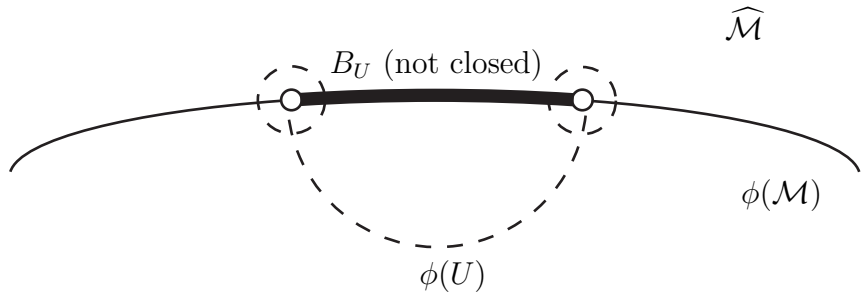}
\caption{the boundary set $B_U$ is not closed because two limit points of $B_U$ (one at each end) are not strongly attached to $U$}
\label{B not closed}
\end{figure}

\begin{figure}[htb!]
\centering%
\psfrag{manifold}{$\widehat{\mathcal{M}}$}
\psfrag{e}{$\phi(\mathcal{M})$}
\psfrag{B}{$B_U$ (closed)}
\psfrag{U}{$\phi(U)$}
\includegraphics{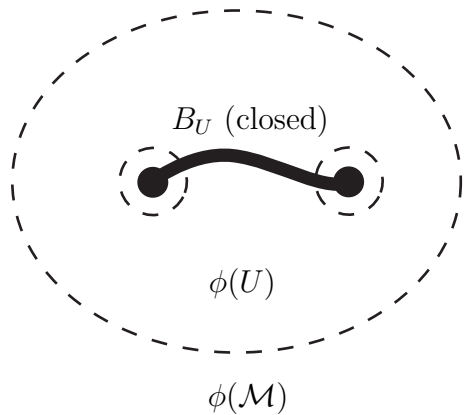}
\caption{the boundary set $B_U$ is closed because all limit points of $B_U$ are strongly attached to $U$}
\label{B closed}
\end{figure}

In general, $B_U$ will not be closed in $\partial(\phi(\mathcal{M}))$ or $\widehat{\mathcal{M}}$ because not all the limit points of $B_U$ are necessarily strongly attached to $U$. See figure \ref{B not closed} and figure \ref{B closed}. As was shown in proposition $13$ and proposition $14$ of \cite{Barry}, however, the limit points of $B_U$ are always attached to $U$.

\section{The Strongly Attached Point Topology}\label{strongly attached topology section}

Similarly to the attached point topology, a basis for a topology on $\overline{\mathcal{M}}=\mathcal{M}\cup\mathcal{B}(\mathcal{M})$ may be constructed by defining the open sets in terms of the strongly attached abstract boundary point definition (definition \ref{strongly attached abstract}).  Again, in keeping with the notion of constructing a natural topology, the open sets of $\mathcal{M}$ to which the abstract boundary points are strongly attached are therefore taken to be the open sets of the manifold topology.

Consider the sets $A_i=U_i\cup B_i$, where $U_i$ is a non-empty open set of the manifold topology in $\mathcal{M}$ and $B_i$ is the set of all abstract boundary points which are strongly attached to $U_i$.  $B_i$ may be the empty set if no abstract boundary points are strongly attached to $U_i$. Let $\mathcal{W}$ be the set comprised of every $A_i$ set.  That is, \begin{displaymath}\mathcal{W}=\{A_i=U_i\cup B_i\}.\end{displaymath}

Given that the strongly attached boundary point definition limits the `type' of open set $U_i$ in $\mathcal{M}$ to which a boundary point may be strongly attached, it is important to know whether or not every abstract boundary point is strongly attached to an open set $U_i$ in $\mathcal{M}$.  In other words, if a boundary point $p$ were attached to a `wedge' shaped open set $U$ in $\mathcal{M}$ like that depicted in figure \ref{attached point}, does there always exist another open set $U'$ in $\mathcal{M}$ to which $p$ is strongly attached?  A brief analysis reveals that the answer to this question is yes - every boundary point $p$ is strongly attached to an open set $U$ in $\mathcal{M}$.

\begin{lemma}\label{all strongly attached}Every abstract boundary point $[p]$ is strongly attached to an open set $U$ in $\mathcal{M}$.\end{lemma}

\textit{Proof:} For the envelopment $\phi:\mathcal{M}\rightarrow\widehat{\mathcal{M}}$, $p\in\partial(\phi(\mathcal{M}))$, let $N$ be any open neighbourhood of $p$ in $\widehat{\mathcal{M}}$.  Since $p$ is a boundary point, $N\cap\phi(\mathcal{M})$ is non-empty.  In addition, since $\phi$ is an embedding, the non-empty set $U=\phi^{-1}(N\cap\phi(\mathcal{M}))$ is an open set in $\mathcal{M}$.  We then have that $p$ is strongly attached to $U=\phi^{-1}(N\cap\phi(\mathcal{M}))$ because there exists an open neighbourhood $N$ of $p$ in $\widehat{\mathcal{M}}$ such that $N\cap\phi(\mathcal{M})\subseteq N\cap\phi(\mathcal{M})$.  $\Box$

\begin{proposition}\label{strongly attached basis}The elements of $\mathcal{W}$ form a basis for a topology on $\overline{\mathcal{M}}$.\end{proposition}

\textit{Proof:} By definition, $\mathcal{M}$ is covered by the collection $\{U_i\}$ of open sets in $\mathcal{M}$.  Also, by lemma \ref{all strongly attached} each abstract boundary point is strongly attached to an open set $\mathcal{U}_i$ in $\mathcal{M}$.  The set of open sets in $\mathcal{M}$ and their strongly attached abstract boundary points, i.e., $\{A_i\}$, therefore covers $\overline{\mathcal{M}}$.

The intersection between two elements of $\mathcal{W}$ must be examined.  Consider the intersection between $A_1=U_1\cup B_1$ and $A_2=U_2\cup B_2$.  In considering this intersection, there are three subcases to check:

\begin{enumerate}\item $U_1\cap U_2 \neq\emptyset$, $B_1\cap B_2=\emptyset$ (this includes the cases when $B_1=\emptyset$ or $B_2=\emptyset$)
\item $U_1\cap U_2\neq\emptyset$, $B_1\cap B_2\neq\emptyset$
\item $U_1\cap U_2=\emptyset$, $B_1\cap B_2\neq\emptyset$\end{enumerate}

$i)$ In the first case we have that $U_1\cap U_2 \neq\emptyset$ and $B_1\cap B_2=\emptyset$.  Since $B_1\cap B_2=\emptyset$, $A_1\cap A_2=U_1\cap U_2=U_3$ which is an open set in $\mathcal{M}$.  If the abstract boundary point $[p]$ is strongly attached to $U_3$, then $[p]$ is strongly attached to $U_1$ ($[p]\in B_1$) and $[p]$ is strongly attached to $U_2$ ($[p]\in B_2$) which would imply that $B_1\cap B_2\neq\emptyset$.  It follows that $B_3=\emptyset$, where $B_3$ is the set of abstract boundary points that are strongly attached to $U_3$, and thus $A_1\cap A_2=U_3\cup B_3\in\mathcal{W}$.
\\

$ii)$ For this case, $A_1\cap A_2=(U_1\cap U_2)\cup(B_1\cap B_2)=U_3\cup(B_1\cap B_2)$. If $[p]\in B_1\cap B_2$, it is strongly attached to both $U_1$ and $U_2$, so by lemma \ref{strong intersection}, $[p]$ is strongly attached to $U_1\cap U_2= U_3$, i.e., $[p]\in B_3$. Thus $B_1\cap B_2\subseteq B_3$. Now if $[p]\in B_3$, it is strongly attached to $U_3=U_1\cap U_2$ and so is strongly attached to both $U_1$ and $U_2$. That is, $[p]\in B_1\cap B_2$ and so $B_3\subseteq B_1\cap B_2$. It follows that $A_1\cap A_2= U_3\cup B_3\in\mathcal{W}$.
\\

$iii)$ This case cannot exist by lemma \ref{strong intersection}.  Specifically, if $B_1\cap B_2\neq\emptyset$, then there exist abstract boundary points that are strongly attached to both $U_1$ and $U_2$, and hence $U_1\cap U_2\neq\emptyset$.
\\

The intersection $A_1\cap A_2=(U_1\cap U_2)\cup(B_1\cap B_2)$ is therefore always another element of $\mathcal{W}$. Thus the elements of $\mathcal{W}$ form a basis for a topology on $\overline{\mathcal{M}}$. $\Box$

\begin{definition}[Strongly attached point topology]The \textit{strongly attached point topology} on $\overline{\mathcal{M}}$ is the topology which has the basis $\mathcal{W}$.\end{definition}

The attached point topology required that sets of abstract boundary points be added to the collection of basis sets $\mathcal{V}$.  This was done to ensure that the sets of $\mathcal{V}$ did in fact define a basis for a topology on $\overline{\mathcal{M}}$.  Because there exist basis sets $A_i=U_i\cup B_i$ and $A_j=U_j\cup B_j$ of the attached point topology such that $A_i\cap A_j=B_i\cap B_j$, i.e., $U_i\cap U_j=\emptyset$, the $C_i$ sets which are collections of abstract boundary points must also be included in the collection $\mathcal{V}$ of basis sets.  The collection $\mathcal{W}$ of basis sets for the strongly attached point topology, however, does not require the addition of such sets of abstract boundary points.  This is a direct consequence of lemma \ref{strong intersection}.  The non-empty intersection of any two $A_i$ sets will necessarily contain points of $\mathcal{M}$, and therefore it is impossible that a collection of abstract boundary points can be produced by considering intersections of $A_i$ sets.  It is for this reason that the topology is referred to as the strongly attached point topology - the abstract boundary $\mathcal{B}(\mathcal{M})$ is firmly affixed to the manifold $\mathcal{M}$ and has become, topologically speaking, an integral part of the larger space $\overline{\mathcal{M}}$. In other words, any open neighbourhood of an abstract boundary point $[p]$ will necessarily include some part of $\mathcal{M}$.

\section{Open and Closed Sets in the Strongly Attached Point Topology}\label{Open and Closed sets}

The open sets of $\overline{\mathcal{M}}$ consist of arbitrary unions of the elements of $\mathcal{W}$.  As in the case of the attached point topology, it may again seem that an arbitrary open set $(U_i\cup B_i)\cup(U_j\cup B_j)\cup...$ is another basis element $U_k\cup B_k$.  Again, this is not true in general.

\begin{figure}[htb!]
\centering%
\psfrag{p}{$p=(0,0)$}
\psfrag{m}{$\widehat{\mathcal{M}}$}
\psfrag{v}{$\phi(\mathcal{M})$}
\psfrag{x}{$x_1$}
\psfrag{b}{$U_1$}
\includegraphics{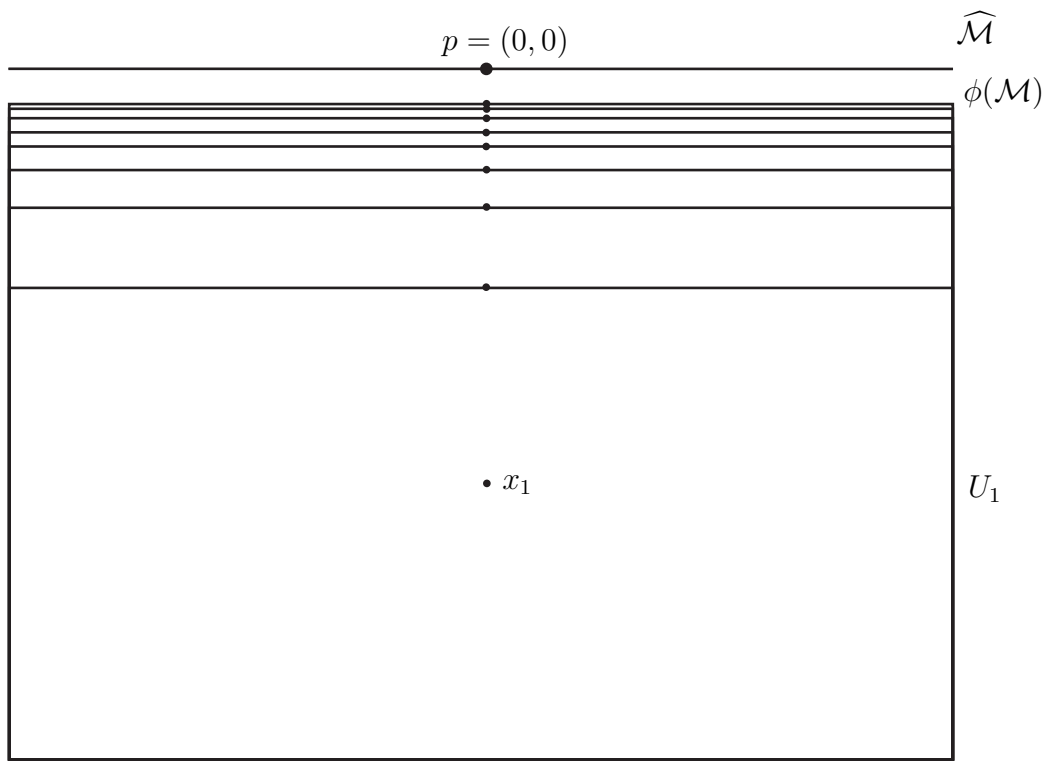}
\caption{the first 8 elements of the sequence $\{x_n\}$ and their open neighbourhoods $U_n$}
\label{counter example}
\end{figure}

\begin{example}Consider $\mathcal{M}=\{(x,y)\in\mathbb{R}^2:y<0\}$, $\widehat{\mathcal{M}}=\mathbb{R}^2$ and let $\phi:\mathcal{M}\rightarrow\widehat{\mathcal{M}}$ be the inclusion map. Let $p\in\partial\phi(\mathcal{M})$ be the boundary point $(0,0)$; $[p]$ is the associated abstract boundary point. Define a sequence $\{x_n\}$ in $\mathcal{M}$ by $x_n\equiv(0,-\frac{1}{n})$. Around each $x_n$ define an open set $U_n=\{(x,y):-1<x<1,-\frac{3}{2}<y<-\frac{1}{n+1}\}$. See figure \ref{counter example}. By construction, in $\widehat{\mathcal{M}}$, for any $n$, $\overline{U_n}\subset\mathcal{M}$ and thus $U_n$ has no strongly attached abstract boundary points, i.e., $B_n=\emptyset$ and $U_n=U_n\cup B_n=A_n$. Take an open ball $\mathcal{B}_{\frac{1}{2}}(p)$ of radius $\frac{1}{2}$ around $p$ and consider some $a\in\mathcal{B}_{\frac{1}{2}}(p)\cap\mathcal{M}$. Because $\{x_n\}\rightarrow p$, $a$ will be contained in some $U_n$. Since this is true for every $a\in\mathcal{B}_{\frac{1}{2}}(p)\cap\mathcal{M}$, it follows that $\mathcal{B}_{\frac{1}{2}}(p)\cap\mathcal{M}\subset\bigcup_n U_n$. The abstract boundary point $[p]$ is therefore strongly attached to the open set $\bigcup_n U_n=O$ but $O$ is the union of non-empty open sets $U_n$ in $\mathcal{M}$, each of which does not have any strongly attached abstract boundary points, i.e., $O=\bigcup_n U_n=\bigcup_n(U_n\cup B_n)$. Since $O\subset\mathcal{M}$ and $[p]\notin O$, $O\notin\mathcal{W}$.\end{example}

\begin{proposition}\label{i(M) open} The set $\mathcal{M}$ is open, and the set $\mathcal{B}(\mathcal{M})$ is closed in the strongly attached point topology on $\overline{\mathcal{M}}$.\end{proposition}

\textit{Proof:} For a manifold $\mathcal{M}$, there exists a complete metric $d$ on $\mathcal{M}$ such that the topology induced by $d$ agrees with the manifold topology of $\mathcal{M}$ \cite{Lorentzian}. Choose $\epsilon>0$, and for each $x\in\mathcal{M}$, let $U_x$ be the open ball $U_x=\{y\in\mathcal{M}:d(x,y)<\epsilon\}$. Now consider the envelopment $\phi:\mathcal{M}\rightarrow\widehat{\mathcal{M}}$ and a boundary point $p\in\partial(\phi(\mathcal{M}))$. We know that $p\notin\overline{\phi(U_x)}$ since $d$ is a complete metric on $\mathcal{M}$ and so $\overline{\phi(U_x)}\subset\phi(\mathcal{M})$.  Thus the set $\widehat{\mathcal{M}}\backslash\overline{\phi(U_x)}$ is an open neighbourhood of $p$ in $\widehat{\mathcal{M}}$ which does not intersect $\phi(U_x)$, and so $p$ is not attached to $U_x$. By lemma \ref{strong=attached point}, $p$ is also not strongly attached to $U_x$. It follows that no boundary point $p$ of any envelopment of $\mathcal{M}$ is strongly attached to $U_x$, which implies that $U_x$ has no strongly attached abstract boundary points, i.e., $B_x=\emptyset$.

Now \begin{eqnarray}\bigcup_{x\in\mathcal{M}} A_x&=&\bigcup_{x\in\mathcal{M}}(U_x\cup B_x)\nonumber\\
&=&(\bigcup_{x\in\mathcal{M}}U_x)\cup(\bigcup_{x\in\mathcal{M}}B_x)\nonumber\\
&=&\mathcal{M}\cup\emptyset=\mathcal{M}\nonumber.\end{eqnarray}

It follows that $\mathcal{M}$ is open in $\overline{\mathcal{M}}$ and thus $B(\mathcal{M})$ is closed because $\overline{\mathcal{M}}\backslash\mathcal{B}(\mathcal{M})=\mathcal{M}$ is open. $\Box$

\begin{proposition}\label{B not open}The set $\mathcal{B}(\mathcal{M})$ is not open, and the set $\mathcal{M}$ is not closed in the strongly attached point topology on $\overline{\mathcal{M}}$.\end{proposition}

\textit{Proof:} Consider any open neighbourhood of an abstract boundary point $[p]\in\mathcal{B}(\mathcal{M})$ in $\overline{\mathcal{M}}$.  Every open set of $\overline{\mathcal{M}}$ is a union of basis sets.  Because every basis set contains a non-empty open subset of $\mathcal{M}$, every open set in the strongly attached point topology will contain a non-empty open subset of $\mathcal{M}$.  Any open set that contains $[p]\in\mathcal{B}(\mathcal{M})$ will therefore necessarily contain some open subset of $\mathcal{M}$ as well, and thus $\mathcal{B}(\mathcal{M})$ cannot be open. Since $\mathcal{B}(\mathcal{M})=\overline{\mathcal{M}}\backslash\mathcal{M}$ is not open, $\mathcal{M}$ is not closed.  $\Box$

\begin{proposition}\label{topologies agree}The manifold topology on $\mathcal{M}$ and the topology induced on $\mathcal{M}$ by the strongly attached point topology on $\overline{\mathcal{M}}$ are the same.\end{proposition}

\textit{Proof:} Let $U\neq\emptyset$ be an open set in $\mathcal{M}$ in the manifold topology. The set $A=U\cup B$, where $B$ is the set of all abstract boundary points which are strongly attached to $U$, is an open set in the strongly attached point topology of $\overline{\mathcal{M}}$. Now $U=A\cap\mathcal{M}$ and thus $U$ is an open set in the topology induced on $\mathcal{M}$ by the strongly attached point topology on $\overline{\mathcal{M}}$. Now let $U\neq\emptyset$ be an open set in the topology induced on $\mathcal{M}$ by the strongly attached point topology on $\overline{\mathcal{M}}$. So $U=V\cap\mathcal{M}$ where $V$ is an open set in the strongly attached point topology on $\overline{\mathcal{M}}$. The set $V$ can be expressed as a union of elements of $\mathcal{W}$, i.e., $V=\bigcup_{i\in I}(U_i\cup B_i)$. Thus $U=(\bigcup_{i\in I}(U_i\cup B_i))\cap\mathcal{M}=(\bigcup_{i\in I}U_i)\cap\mathcal{M}=\bigcup_{i\in I} U_i$, and so $U$ is a union of non-empty open sets of the manifold topology on $\mathcal{M}$ and is, therefore, itself an open set of the manifold topology on $\mathcal{M}$. $\Box$

\begin{corollary}\label{non empty intersection}If $V$ is an open neighbourhood of the abstract boundary point $[p]$ in $\overline{\mathcal{M}}$, then $V\cap\mathcal{M}\neq\emptyset$.\end{corollary}

\textit{Proof:} This result follows immediately from the proof of proposition \ref{B not open}. $\Box$
\\

Proposition \ref{B not open} and corollary \ref{non empty intersection} summarise the key difference between the attached point topology and the strongly attached point topology. In contrast to the attached point topology, it is not possible to construct an open set of abstract boundary points in the strongly attached point topology that does not also intersect $\mathcal{M}$. This is demonstrated by corollary \ref{non empty intersection}. It was also previously shown that the set of abstract boundary points is both open and closed with respect to the attached point topology, thus resulting in the set of all abstract boundary points being disconnected from $\mathcal{M}$. Proposition \ref{B not open} shows that this is not the case for the strongly attached point topology. As discussed in the introduction, the abstract boundary points can therefore be regarded as being firmly affixed to the manifold $\mathcal{M}$ with respect to the strongly attached point topology.

\begin{corollary}\label{open set in m}If $V$ is an open neighbourhood of the abstract boundary point $[p]$ in $\overline{\mathcal{M}}$, then $[p]$ is strongly attached to the open set $U\neq\emptyset$ of $\mathcal{M}$, where $U\equiv V\cap\mathcal{M}$.\end{corollary}

\textit{Proof:} $V$ is an open neighbourhood of $[p]$ in $\overline{\mathcal{M}}$ and thus it may be written as a union of basis sets $U_i\cup B_i$, where $[p]$ is an element of at least one of the $B_i$ sets. It follows that $U=V\cap\mathcal{M}=\bigcup_i U_i$. By proposition \ref{topologies agree}, we know that $U$ is an open set of $\mathcal{M}$, and by corollary \ref{non empty intersection}, we have that $\bigcup_i U_i\neq\emptyset$. Moreover, $[p]$ is strongly attached to one of the $U_i$ sets and thus $[p]$ is strongly attached to $U$ where $U\neq\emptyset$. $\Box$
\\

We will next consider if the singleton abstract boundary point sets $\{[p]\}$ are open or closed in the strongly attached point topology.  Before addressing that question, however, three useful results are established.

\begin{proposition}\label{cover iff strongly attached}For abstract boundary points $[p]$ and $[q]$, $[p]\triangleright[q]$ if and only if $[q]$ is strongly attached to every open set, $U\subseteq\mathcal{M}$, to which $[p]$ is strongly attached.\end{proposition}

\textit{Proof:} ($\Leftarrow$) Suppose that $[q]$ is strongly attached to every open set, $U\subseteq\mathcal{M}$, to which $[p]$ is strongly attached. Consider the boundary point $p$ of the envelopment $\phi:\mathcal{M}\rightarrow\widehat{\mathcal{M}}$, and the boundary point $q$ of the envelopment $\phi':\mathcal{M}\rightarrow\widehat{\mathcal{M}'}$. Let $N$ be an open neighbourhood of $p$ in $\widehat{\mathcal{M}}$.  It follows that $[p]$ is strongly attached to $\phi^{-1}(N\cap\phi(\mathcal{M}))$.  Because $[q]$ is strongly attached to every open set $U\subseteq\mathcal{M}$ to which $[p]$ is strongly attached, $[q]$ is strongly attached to $\phi^{-1}(N\cap\phi(\mathcal{M}))$, for all $N$.  For every $N$ there therefore exists an open neighbourhood $W$ of $q$ in $\widehat{\mathcal{M}'}$ such that $W\cap\phi'(\mathcal{M})\subseteq\phi'\circ\phi^{-1}(N\cap\phi(\mathcal{M}))$, i.e., $\{p\}$ covers $\{q\}$, and thus $[p]\triangleright[q]$.
\\

($\Rightarrow$) This follows immediately from proposition \ref{B strongly attached}.  $\Box$

\begin{corollary}\label{limit point}For abstract boundary points $[p]$ and $[q]$, $[p]\triangleright[q]$ if and only if every open neighbourhood of $[p]$ in $\overline{\mathcal{M}}$ also contains $[q]$, where $\overline{\mathcal{M}}$ has the strongly attached point topology.\end{corollary}

\textit{Proof:} ($\Leftarrow$) Suppose that every open neighbourhood of $[p]$ in $\overline{\mathcal{M}}$ also contains $[q]$, where $\overline{\mathcal{M}}$ has the strongly attached point topology. Also suppose that $[p]$ is strongly attached to the open set $U\subseteq\mathcal{M}$. Now define the set $V=U\cup B_U$, where $B_U$ is the set of all abstract boundary points which are strongly attached to $U$. It is clear that $V$ is an open set in $\overline{\mathcal{M}}$ in the strongly attached point topology, $[p]\in V$, and thus $V$ is an open neighbourhood of $[p]$ in $\overline{\mathcal{M}}$. By assumption, $[q]\in V$, and therefore $[q]$ is strongly attached to the open set $U\subseteq\mathcal{M}$. It follows from proposition \ref{cover iff strongly attached} that $[p]\triangleright[q]$.
\\

($\Rightarrow$) Suppose now that $[p]\triangleright[q]$. Let $V$ be an open neighbourhood of $[p]$ in $\overline{\mathcal{M}}$, where $\overline{\mathcal{M}}$ has the strongly attached point topology. The set $V$ can be expressed as a union of elements of $\mathcal{W}$, i.e., $V=\bigcup_{i\in I}(U_i\cup B_i)$. The abstract boundary point $[p]$ lies in at least one set $B_i$ and is therefore strongly attached to the open set $U_i$ of $\mathcal{M}$. It follows from proposition \ref{cover iff strongly attached} that $[q]$ is also strongly attached to $U_i$, and so $[q]\in B_i$. Thus $V$ is also an open neighbourhood of $[q]$. $\Box$

\begin{corollary}\label{abstract closure}The closure in $\overline{\mathcal{M}}$ of an abstract boundary point $[p]$ is $\overline{\{[p]\}}=\{[p]\}\cup\{[x]:[x]\in\mathcal{B}(\mathcal{M}), [x]\triangleright[p]\}$.\end{corollary}

\textit{Proof:} From proposition \ref{i(M) open}, $\mathcal{B}(\mathcal{M})$ is closed in the strongly attached point topology on $\overline{\mathcal{M}}$. Since $[p]\in\mathcal{B}(\mathcal{M})$, it follows that $\overline{\{[p]\}}\subseteq\mathcal{B}(\mathcal{M})$. Now consider $[x]\in\mathcal{B}(\mathcal{M})$.

\begin{tabular}{llp{10cm}}
$[x]\in\overline{\{[p]\}}$&$\Leftrightarrow$&every closed subset of $\mathcal{B}(\mathcal{M})$ that contains $[p]$ also contains $[x]$\\
\phantom{a}&$\Leftrightarrow$&there exists no closed subset of $\mathcal{B}(\mathcal{M})$ that contains $[p]$ that does not contain $[x]$\\
\phantom{a}&$\Leftrightarrow$&there exists no open neighbourhood of $[x]$ in $\overline{\mathcal{M}}$ that does not contain $[p]$\\
\phantom{a}&$\Leftrightarrow$&every open neighbourhood of $[x]$ in $\overline{\mathcal{M}}$ contains $[p]$\\
\phantom{a}&$\Leftrightarrow$&$[x]\triangleright[p]$ (by proposition \ref{limit point})
\end{tabular}

Thus $\overline{\{[p]\}}=\{[p]\}\cup\{[x]:[x]\in\mathcal{B}(\mathcal{M}),[x]\triangleright[p]\}$. $\Box$
\\

We may now more readily consider the question of whether or not the singleton abstract boundary point sets $\{[p]\}$ are open or closed in the strongly attached point topology on $\overline{\mathcal{M}}$.

\begin{proposition}The singleton abstract boundary point sets $\{[p]\}$ are not open in the strongly attached point topology on $\overline{\mathcal{M}}$. They are also not closed in the strongly attached point topology on $\overline{\mathcal{M}}$ if and only if there exists $[q]\in\mathcal{B}(\mathcal{M})$, $[q]\neq[p]$, such that $[q]\triangleright[p]$.\end{proposition}

\textit{Proof:} Consider an abstract boundary point $[p]$. From corollary \ref{non empty intersection}, any open set in $\overline{\mathcal{M}}$ that contains $[p]$ will necessarily have non-empty intersection with $\mathcal{M}$, and thus $\{[p]\}$ is not an open set.

By corollary \ref{abstract closure}, the closure in $\overline{\mathcal{M}}$ of an abstract boundary point $[p]$ is $\overline{\{[p]\}}=\{[p]\}\cup\{[x]:[x]\in\mathcal{B}(\mathcal{M}),[x]\triangleright[p]\}$. If there exists a $[q]$ such that $[q]\triangleright[p]$, $[q]\neq[p]$, then by corollary \ref{abstract closure}, $\overline{\{[p]\}}$ contains at least $[p]$ and $[q]$, and therefore $\{[p]\}$ is not closed. Similarly, if $\{[p]\}$ is not closed, then $\overline{\{[p]\}}$ contains at least $2$ elements $[p]$ and $[q]$ such that $[p]\neq[q]$ and $[q]\triangleright[p]$. It follows that $\{[p]\}$ is not closed in the strongly attached point topology on $\overline{\mathcal{M}}$ if and only if there exists $[q]\in\mathcal{B}(\mathcal{M})$, $[q]\neq[p]$, such that $[q]\triangleright[p]$. $\Box$

\begin{proposition} The open sets of the induced topology on $\mathcal{B}(\mathcal{M})\subset\overline{\mathcal{M}}$, where $\overline{\mathcal{M}}$ has the strongly attached point topology, are arbitrary unions of the $B_i$ sets defined in the basis $\mathcal{W}$.\end{proposition}

\textit{Proof:} Let $\mathcal{T}_{\overline{\mathcal{M}}}$ be the strongly attached point topology on $\overline{\mathcal{M}}$.  The subspace topology on $\mathcal{B}(\mathcal{M})$ is the collection of sets $\mathcal{T}_{\mathcal{B}(\mathcal{M})}=\{U\cap\mathcal{B}(\mathcal{M}):U\in \mathcal{T}_{\overline{\mathcal{M}}}\}$. The topology $\mathcal{T}_{\overline{\mathcal{M}}}$ is the collection of arbitrary unions of the $U_i\cup B_i$ sets of the basis $\mathcal{W}$.  The intersection of these sets with $\mathcal{B}(\mathcal{M})$ is therefore the collection of arbitrary unions of the $B_i$ sets. $\Box$

\section{The Inclusion Map from $\mathcal{M}$ to $\overline{\mathcal{M}}$}\label{the inclusion map}

We now consider the inclusion map $i:\mathcal{M}\rightarrow\overline{\mathcal{M}}=\mathcal{M}\cup\mathcal{B}(\mathcal{M})|i(p)=p$. As in the case of the attached point topology, it can be shown that the inclusion map is an embedding.

\begin{proposition} If $\overline{\mathcal{M}}$ has the strongly attached point topology, then the inclusion mapping $i:\mathcal{M}\rightarrow\overline{\mathcal{M}}|i(p)=p$ is an embedding.\end{proposition}

\textit{Proof:} The inclusion mapping $i$ is an embedding if it is a homeomorphism of $\mathcal{M}$ onto $i(\mathcal{M})$ in the subspace topology on $i(\mathcal{M})\cap\overline{\mathcal{M}}$.  Clearly $i$ is a bijection of $\mathcal{M}$ onto $i(\mathcal{M})$.  Now let $\mathcal{T}_\mathcal{M}$ be the usual topology on $\mathcal{M}$ consisting of the collection of open sets $\{U_i\}$, $\mathcal{T}_{\overline{\mathcal{M}}}$ the strongly attached point topology on $\overline{\mathcal{M}}$ as defined by the basis elements of $\mathcal{W}$, i.e., $\mathcal{T}_{\overline{\mathcal{M}}}$ is the collection of arbitrary unions of the $U_i\cup B_i$ sets, and $\mathcal{T}_{i(\mathcal{M})}$ the subspace topology on $i(\mathcal{M})\cap\overline{\mathcal{M}}$.  The subspace topology $\mathcal{T}_{i(\mathcal{M})}$ is therefore the collection of sets $\mathcal{T}_{i(\mathcal{M})}=\{U_k\}$.  Clearly both $i$ and $i^{-1}$ are continuous with respect to $\mathcal{T}_{\mathcal{M}}$ and $\mathcal{T}_{i(\mathcal{M})}$.  It has thus been demonstrated that $i:\mathcal{M}\rightarrow\overline{\mathcal{M}}\mid i(p)=p$ is a homeomorphism onto its image in the induced topology and is therefore an embedding. $\Box$
\\

Because it has been shown that $i:\mathcal{M}\rightarrow\overline{\mathcal{M}}\mid i(p)=p$ is an embedding, we may view $\overline{\mathcal{M}}$ as simply $\mathcal{M}$ with the addition of its abstract boundary points.  This is a pleasing result as one would expect the nature of $\mathcal{M}$ to be preserved in $\overline{\mathcal{M}}$.

The following properties of $i(\mathcal{M})$ are readily obtained.

\begin{proposition}For the inclusion mapping $i:\mathcal{M}\rightarrow\overline{\mathcal{M}}|i(p)=p$, $i(\mathcal{M})$ is open and not closed in the strongly attached point topology on $\overline{\mathcal{M}}$, $\overline{i(\mathcal{M})}=\overline{\mathcal{M}}$ and $\partial(i(\mathcal{M}))=\mathcal{B}(\mathcal{M})$.\end{proposition}

\textit{Proof:} Since $i(\mathcal{M})=\mathcal{M}$, it follows from proposition \ref{i(M) open} and proposition \ref{B not open} that $i(\mathcal{M})$ is open and not closed in the strongly attached point topology on $\overline{\mathcal{M}}$. Because $i(\mathcal{M})$ is open, $\partial(i(\mathcal{M}))=\partial(\mathcal{M})=\{x\in\overline{\mathcal{M}}\backslash\mathcal{M}:\textrm{every open neighbourhood of } x \textrm{ has non-empty intersection with } \mathcal{M}\}$. Consider an abstract boundary point $[p]\in\mathcal{B}(\mathcal{M})=\overline{\mathcal{M}}\backslash\mathcal{M}$. From corollary \ref{non empty intersection}, every open neighbourhood of $[p]$ has non-empty intersection with $\mathcal{M}$, and so $[p]\in\partial(i(\mathcal{M}))$. Thus $\partial(i(\mathcal{M}))=\mathcal{B}(\mathcal{M})$. Now $\overline{i(\mathcal{M})}=i(\mathcal{M})\cup\partial(i(\mathcal{M}))=\mathcal{M}\cup\mathcal{B}(\mathcal{M})=\overline{\mathcal{M}}$. $\Box$

\section{Contact Properties of the Strongly Attached Point Topology}\label{Contact properties}

A number of important properties of the strongly attached point topology will now be presented.

Due to the way that abstract boundary points are constructed, two abstract boundary points may share some of the same topological information.  For example, if $[p]=[q]$ then any envelopment that produces a boundary set belonging to $[p]$ will also produce a boundary set belonging to $[q]$ and vice versa.  Likewise, in the case that $[p]$ covers $[q]$ we have that $[p]$ contains $[q]$ in some sense. When $[p]$ and $[q]$ are realised as boundary sets $A\subseteq\partial(\phi(\mathcal{M}))$ and $B\subseteq\partial(\phi'(\mathcal{M}))$ respectively, the topological structure of $\phi(\mathcal{M})$ near $A$ incorporates the topological structure of $\phi'(\mathcal{M})$ near $B$. In this way, when we consider the abstract boundary point $[p]$ relative to $\mathcal{M}$, we are also considering the abstract boundary point $[q]$. Alternatively, we may have the case where $[p]$ and $[q]$ are not in contact at all, and are somehow `separate' from each other.

A topology on $\overline{\mathcal{M}}$ should therefore be descriptive of the topological `contact' properties between abstract boundary points.  It can be seen that the strongly attached point topology describes the separation properties of abstract boundary points in a natural way in that greater levels of separation between abstract boundary points with respect to the covering relation correspond to greater levels of separation with respect to the usual topological separation axioms.

We begin by defining what it means for an abstract boundary point to be in contact with another abstract boundary point.  In some sense the contact relation is a weaker form of the covering relation.  If $[p]$ and $[q]$ are in contact, then they contain some of the same topological information, but not as much as if $[p]$ covered $[q]$ or $[q]$ covered $[p]$.

\begin{definition}[Contact $\perp$]\label{neighbourhood contact defn} Let $p\in\partial(\phi(\mathcal{M}))$ and $q\in\partial(\phi'(\mathcal{M}))$ be two enveloped boundary points of $\mathcal{M}$. They are said to be \textit{in contact} (denoted $p\perp q$) if for all open neighbourhoods $U$ and $V$ of $p$ and $q$ respectively, $$U\sqcap V:=\phi^{-1}(U\cap\phi
(\mathcal{M}))\cap\phi'^{-1}(V\cap\phi'(\mathcal{M}))\neq\emptyset.$$ \end{definition}

\begin{definition}[Contact $\perp$ (sequence definition)]\label{sequence contact defn} Two boundary points $p\in\partial(\phi(\mathcal{M}))$ and $q\in\partial(\phi'(\mathcal{M}))$ are \textit{in contact} (denoted $p\perp q$) if there exists a sequence $\{p_i\}\subset\mathcal{M}$ such that $\{\phi(p_i)\}$ has $p$ as an endpoint and $\{\phi'(p_i)\}$ has $q$ as an endpoint.\end{definition}

Definitions \ref{neighbourhood contact defn} and \ref{sequence contact defn} are equivalent. For a proof of this see lemma 6.3 of \cite{Ashley}.

\begin{definition}[Abstract boundary points in contact]\label{abstract contact} Two abstract boundary points $[p]$ and $[q]$ are \textit{in contact}, denoted $[p]\perp[q]$, if $p\perp q$ for boundary point representatives $p$ and $q$. \end{definition}

This definition can be shown to be well-defined. See theorem 3.10 of \cite{Ashley}.

\begin{definition}[Separation of boundary points $\parallel$] Two boundary points $p\in\partial(\phi(\mathcal{M}))$ and $q\in\partial(\phi'(\mathcal{M}))$ are \textit{separate} (denoted $p\parallel q$) if there is no sequence $\{p_i\}\subset\mathcal{M}$ for which $\{\phi(p_i)\}\rightarrow p$ and $\{\phi'(p_i)\}\rightarrow q$. Equivalently, the boundary points $p$ and $q$ are separate if there exist open neighbourhoods $U$ and $V$ of $p$ and $q$ respectively such that $\phi^{-1}(U\cap\phi
(\mathcal{M}))\cap\phi'^{-1}(V\cap\phi'(\mathcal{M}))=\emptyset$.\end{definition}

Equivalently, from definition \ref{sequence contact defn}, two boundary points $p\in\partial(\phi(\mathcal{M}))$ and $q\in\partial(\phi'(\mathcal{M}))$ are separate if they are not in contact.

\begin{definition}[Separation of abstract boundary points]\label{abstract seperate} Two abstract boundary points $[p]$ and $[q]$ are \textit{separate}, denoted $[p]\parallel[q]$, if $p\parallel q$ for boundary point representatives $p$ and $q$.  Equivalently, $[p]\parallel[q]$ if they are not in contact.\end{definition}

Similar to definition \ref{abstract contact}, this definition can be shown to be well-defined. See theorem 3.10 of \cite{Ashley}.

The results which follow relate to $\overline{\mathcal{M}}$ with the strongly attached point topology.

\begin{proposition}\label{hausdorff separated} Two abstract boundary points $[p]$ and $[q]$ are $T_2$ separable, i.e., they are Hausdorff separable, if and only if $[p]\parallel[q]$.\end{proposition}

\textit{Proof:} $(\Leftarrow)$ If $[p]\parallel[q]$, then $[p]$ and $[q]$ are $T_2$ separable.
\\

If $[p]\parallel[q]$ then there exists an open neighbourhood $U$ of $p\in\partial(\phi(\mathcal{M}))$ and an open neighbourhood $V$ of $q\in\partial(\phi'(\mathcal{M}))$ such that $\phi^{-1}(U\cap\phi(\mathcal{M}))\cap\phi'^{-1}(V\cap\phi'(\mathcal{M}))=\emptyset$.  We also have that $p$ is strongly attached to $\phi^{-1}(U\cap\phi(\mathcal{M}))$, $q$ is strongly attached to $\phi'^{-1}(V\cap\phi'(\mathcal{M}))$, and from lemma \ref{strong intersection}, $p$ is not strongly attached to $\phi'^{-1}(V\cap\phi'(\mathcal{M}))$ and $q$ is not strongly attached to $\phi^{-1}(U\cap\phi(\mathcal{M}))$. Furthermore, it also follows from lemma \ref{strong intersection} that there exist no abstract boundary points which are strongly attached to both $\phi^{-1}(U\cap\phi(\mathcal{M}))$ and $\phi'^{-1}(V\cap\phi'(\mathcal{M}))$. There therefore exists an open neighbourhood of $[p]$, $\phi^{-1}(U\cap\phi(\mathcal{M}))\cup B_U$, $[p]\in B_U$, and an open neighbourhood of $[q]$, $\phi'^{-1}(V\cap\phi'(\mathcal{M}))\cup B_V$, $[q]\in B_V$, such that their intersection is empty.  The abstract boundary points $[p]$ and $[q]$ are therefore Hausdorff separated.
\\

$(\Rightarrow)$ If $[p]$ and $[q]$ are $T_2$ separable, then $[p]\parallel[q]$.
\\

If $[p]$ and $[q]$ are Hausdorff separated then there exist open neighbourhoods $U_p$ of $[p]$ and $U_q$ of $[q]$ in the strongly attached point topology such that $U_p\cap U_q=\emptyset$.  Since $[p]$ is contained in $U_p$, by corollary \ref{open set in m}, $[p]$ is strongly attached to the open set $V_p=U_p\cap\mathcal{M}$.  Now, since $[p]$ is strongly attached to $V_p$, there exists an open neighbourhood $W_p$ in $\widehat{\mathcal{M}}$ of $p\in\partial(\phi(\mathcal{M}))$ such that $W_p\cap\phi(\mathcal{M})\subseteq\phi(V_p)$. Similarly, there exists an open neighbourhood $W_q$ in $\widehat{\mathcal{M}'}$ of $q\in\partial(\phi'(\mathcal{M}))$ such that $W_q\cap\phi'(\mathcal{M})\subseteq\phi'(V_q)$, where $V_q$ is the open set $V_q=U_q\cap\mathcal{M}$ in $\mathcal{M}$. Now, since $W_p\cap\phi(\mathcal{M})\subseteq\phi(V_p)$ and $W_q\cap\phi'(\mathcal{M})\subseteq\phi'(V_q)$, and $V_p\subseteq U_p$ and $V_q\subseteq U_q$, where $U_p\cap U_q=\emptyset$, it follows that $\phi^{-1}(W_p\cap\phi(\mathcal{M}))\cap\phi'^{-1}(W_q\cap\phi'(\mathcal{M}))=\emptyset$, and so $[p]\parallel[q]$. $\Box$

\begin{proposition}\label{T_1 separated}Two abstract boundary points $[p]$ and $[q]$ are $T_1$ separated if and only if $[p]\ntriangleright[q]$ and $[q]\ntriangleright[p]$.\end{proposition}

\textit{Proof:} ($\Leftarrow$) If $[p]\ntriangleright[q]$ and $[q]\ntriangleright[p]$, then $[p]$ and $[q]$ are $T_1$ separated.
\\

If $[p]\ntriangleright[q]$ and $[q]\ntriangleright[p]$, then by corollary \ref{limit point} there exists an open neighbourhood $N_p$ of $[p]$ and an open neighbourhood $N_q$ of $[q]$ such that $[q]\notin N_p$ and $[p]\notin N_q$.  This is a statement of the $T_1$ separation axiom.
\\

($\Rightarrow$) If $[p]$ and $[q]$ are $T_1$ separated, then $[p]\ntriangleright[q]$ and $[q]\ntriangleright[p]$.
\\

If $[p]$ and $[q]$ are $T_1$ separated, then there exists an open neighbourhood $N_p$ of $[p]$ and an open neighbourhood $N_q$ of $[q]$ such that $[q]\notin N_p$ and $[p]\notin N_q$.  It then follows directly from corollary \ref{limit point}, that $[p]\ntriangleright[q]$ and $[q]\ntriangleright[p]$. $\Box$

\begin{proposition}\label{T0 sep1}Two abstract boundary points $[p]$ and $[q]$ are $T_0$ separated if and only if $[p]\ntriangleright[q]$ or $[q]\ntriangleright[p]$.\end{proposition}

\textit{Proof:} ($\Leftarrow$) If $[p]\ntriangleright[q]$ or $[q]\ntriangleright[p]$, then $[p]$ and $[q]$ are $T_0$ separated.
\\

By corollary \ref{limit point}, if $[q]\ntriangleright[p]$, then there exists an open neighbourhood $N_q$ of $[q]$ such that $[p]\notin N_q$, and so $[p]$ and $[q]$ are $T_0$ separated. Likewise, if $[p]\ntriangleright[q]$, then $[p]$ and $[q]$ are $T_0$ separated.
\\

($\Rightarrow$) If $[p]$ and $[q]$ are $T_0$ separated, then $[p]\ntriangleright[q]$ or $[q]\ntriangleright[p]$.
\\

If $[p]$ and $[q]$ are $T_0$ separated, then there exists an open neighbourhood $N_p$ of $[p]$ such that $[q]\notin N_p$, or there exists an open neighbourhood $N_q$ of $[q]$ such that $[p]\notin N_q$.  If $[p]\notin N_q$, then by corollary \ref{limit point}, $[q]\ntriangleright[p]$.  Likewise, if $q\notin N_p$, then $[p]\ntriangleright[q]$. $\Box$
\\

\begin{figure}[htb!]
\centering%
\psfrag{r}{$\mathbb{R}^2$}
\psfrag{z}{$\mathbb{R}^2$}
\psfrag{p}{$p$}
\psfrag{b}{$B$}
\psfrag{c}{$\phi(\mathcal{M})$}
\psfrag{d}{$\phi'(\mathcal{M})$}
\psfrag{q}{$q$}
\includegraphics{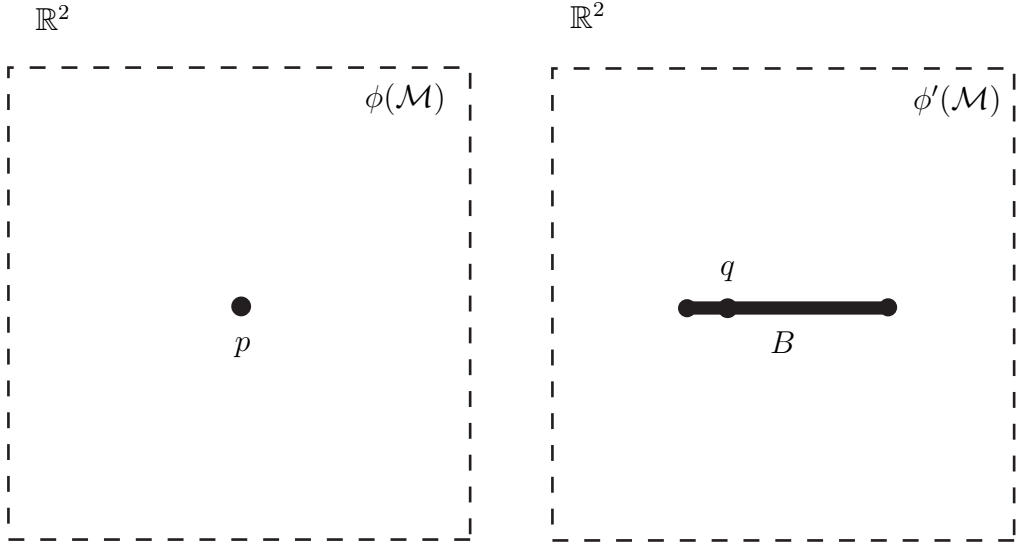}
\caption{the boundary point $p\in\partial(\phi(\mathcal{M}))$ is equivalent to the closed boundary set $B\subset\partial(\phi'(\mathcal{M}))$, where $q\in B$. It follows that $p\triangleright q$, but $q\ntriangleright p$.}
\label{second case}
\end{figure}

\begin{figure}[htb!]
\centering%
\psfrag{a}{$\phi'(\mathcal{M})$}
\psfrag{b}{$\widehat{\mathcal{M}'}$}
\psfrag{c}{$\phi'(\lambda_2)$}
\psfrag{d}{$\phi'(\lambda_1)$}
\psfrag{q}{$q$}
\psfrag{e}{$\phi(\mathcal{M})$}
\psfrag{f}{$\widehat{\mathcal{M}}$}
\psfrag{h}{$\phi(\lambda_2)$}
\psfrag{g}{$\phi(\lambda_1)$}
\psfrag{p}{$p$}
\psfrag{t}{$t$}
\psfrag{z}{$\psi$}
\psfrag{x}{$t'$}
\psfrag{y}{$\psi'$}
\psfrag{j}{$\partial(\phi(\mathcal{M}))$}
\psfrag{k}{$\partial(\phi'(\mathcal{M}))$}
\includegraphics[scale=0.7]{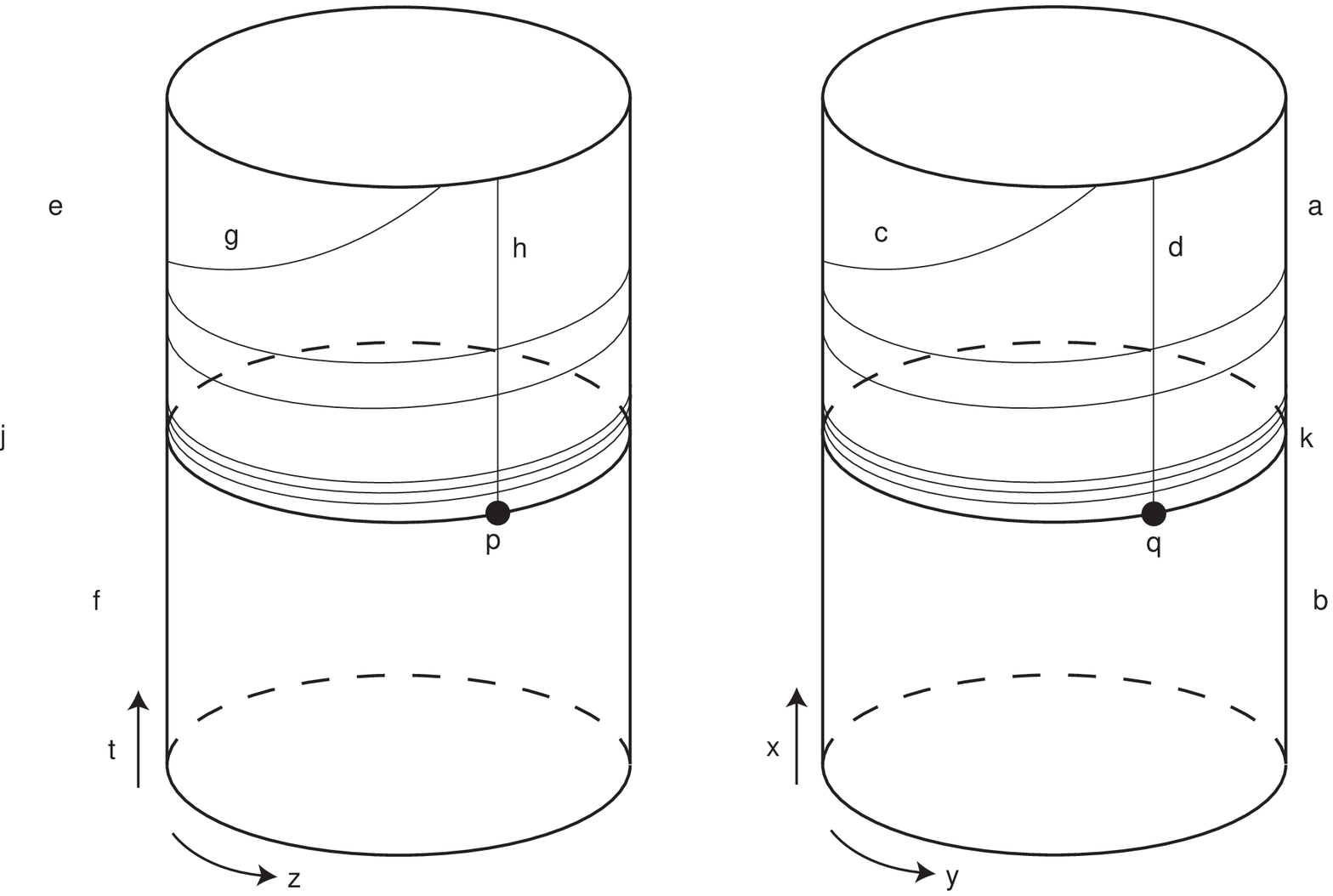}
\caption{two envelopments of the two-dimensional Misner space-time with respective metrics:  $ds^2=2d\psi dt+t(d\psi)^2$ and $ds^2=-2d\psi' dt'+t'(d\psi')^2$. The curves $\lambda_1$ and $\lambda_2$ are null geodesics. We may construct a sequence along $\phi(\lambda_1)$ that converges to $p$. It follows that the image of this sequence under $\phi'$ converges to $q$. The boundary points $p$ and $q$ are therefore in contact. The curve $\phi'(\lambda_2)$ is an element of a class of geodesics that spiral around the space-time and approach the waist. The image under $\phi$ of each such geodesic is a straight vertical line similar to $\phi(\lambda_2)$ that approaches some point of the boundary set $\partial(\phi(\mathcal{M}))$ of which $p$ is an element. It follows that $p\ntriangleright q$ as we can construct a sequence that converges to $q$ along one of the spiraling geodesics in $\phi'(\mathcal{M})$ whose image under $\phi$ does not converge to $p$. By a similar argument it can be shown that $q\ntriangleright p$.}
\label{third case}
\end{figure}

The results of this section are summarised in table 1 which shows the correspondence between the contact properties of two enveloped boundary points $p\in\partial(\phi(\mathcal{M}))$ and $q\in\partial(\phi'(\mathcal{M}))$ and the topological relationship of the respective abstract boundary points $[p]$ and $[q]$ in $\overline{\mathcal{M}}$ with the strongly attached point topology. We provide examples of the second and third relationships in figures \ref{second case} and \ref{third case}, respectively.

\begin{table}[htbp]
\begin{tabular}{|p{7cm}|p{7cm}|}
  \hline
  Relationship \vspace{-0.7em} between enveloped boundary points \vspace{-0.7em} $p\in\partial(\phi(\mathcal{M}))$ and $q\in\partial(\phi'(\mathcal{M}))$ & Topological \vspace{-0.7em} relationship of the abstract boundary points $[p]$ and $[q]$  \\ \hline \hline \hline
  $p\sim q$ & $[p]=[q]$ \\ \hline
  $p\triangleright q$, $q\ntriangleright p$ or \vspace{-0.7em} \newline $q\triangleright p$, $p\ntriangleright q$
  & $[p]$ and $[q]$ are $T_0$ separated \vspace{-0.7em} (proposition \ref{T0 sep1})\newline
  $[p]$ and $[q]$ are not $T_1$ \vspace{-0.7em} separated (proposition \ref{T_1 separated}) \\ \hline
  $p\perp q$, $p\ntriangleright q$, $q\ntriangleright p$ & $[p]$ and $[q]$ are $T_1$ separated \vspace{-0.7em} (proposition \ref{T_1 separated})\newline
  $[p]$ and $[q]$ are not $T_2$ \vspace{-0.7em} separated (proposition \ref{hausdorff separated}) \\ \hline
  $p\parallel q$& $[p]$ and $[q]$ are $T_2$ separated \vspace{-0.7em} (proposition \ref{hausdorff separated})\\
  \hline
\end{tabular}
\caption{The left hand column shows the possible relationships between boundary points $p$ and $q$ of envelopments $\phi$ and $\phi'$, respectively, of $\mathcal{M}$; the right hand column shows the corresponding topological relationships between the associated abstract boundary points $[p]$ and $[q]$ in $\overline{\mathcal{M}}$ with the strongly attached point topology.}\end{table}

Hausdorff separability is lost between abstract boundary points which are in contact with each other, and therefore, also when one of the abstract boundary points covers the other. In many ways, this is an expected result.  As has been stated previously, two abstract boundary points which are in contact with one another share a certain amount of topological information, and thus they do not represent two truly distinct points.  This property is reflected in the loss of Hausdorff separation in the strongly attached point topology. And so, while it is desirable that a topology for $\overline{\mathcal{M}}$ be Hausdorff, it can be seen that the lack of separation between abstract boundary points actually provides us with information about the structure of the abstract boundary itself. Moreover, it can be argued that Hausdorff separation is not lost between truly distinct abstract boundary points (namely those which are separate).  Instead, it is lost between abstract boundary points which represent different parts of some `larger' entity.

We note that, in general, the strongly attached point topology on $\overline{\mathcal{M}}$ will be $T_0$ separated only, as there will be occurrences of $p\triangleright q$, $q\ntriangleright p$ for boundary points $p\in\partial(\phi(\mathcal{M}))$ and $q\in\partial(\phi'(\mathcal{M}))$.
\\

We will now determine if the strongly attached point topology is first countable.

\begin{proposition}The strongly attached point topology on $\overline{\mathcal{M}}$ is first countable.\end{proposition}

\textit{Proof:} A topological space $X$ is said to be \textit{first countable} if, for each $x\in X$, there exists a sequence $U_1$, $U_2$,... of open neighbourhoods of $x$ such that for any open neighbourhood, $V$, of $x$, there exists an integer, $i$, such that $U_i\subseteq V$.

For $X=\overline{\mathcal{M}}$ with the strongly attached point topology, we firstly consider the case where $x\in\mathcal{M}$.  Given the existence of a complete metric $d$ on $\mathcal{M}$, we know from the proof of proposition \ref{i(M) open}, that for $n\in\mathds{N}$, the open balls $U_n=\{p\in\mathcal{M}:d(x,p)<1/n\}$ based at the point $x$ have no attached abstract boundary points and therefore no strongly attached abstract boundary points.  The sets $U_n\cup B_n=U_n$ are basis elements of $\mathcal{W}$, and so $U_1$, $U_2$,... is a sequence of open neighbourhoods of $x$.

Let $V$ be an open neighbourhood of $x$. Thus $V$ is an arbitrary union of basis elements $A_i$ which implies that $x\in A_k=U_k\cup B_k\subseteq V$ for some $A_k$ in the union. It is possible to choose an $n\in\mathds{N}$, such that, for the open ball $U_n$, $\overline{U_n}\subset U_k$.  Thus $U_n\subseteq V$. We have therefore shown that $\overline{\mathcal{M}}$ is first countable at $x$, for all $x\in\mathcal{M}$.

Now we consider an abstract boundary point $[p]\in\mathcal{B}(\mathcal{M})$, where $p$ is a boundary point of some envelopment $(\mathcal{M},\widehat{\mathcal{M}},\phi)$. Similarly to before, given the existence of a complete metric $d$ on $\widehat{\mathcal{M}}$, we can define a series of open balls of $p$ in $\widehat{\mathcal{M}}$ by $O_n=\{y\in\widehat{\mathcal{M}}:d(p,y)<1/n\}$, $n\in\mathds{N}$.  We therefore have a series of sets in $\overline{\mathcal{M}}$ that contain $[p]$ defined by $[\phi^{-1}(O_n\cap\phi(\mathcal{M}))]\cup B_{O_n}$, where the $B_{O_n}$ are the collections of abstract boundary points that are strongly attached to $\phi^{-1}(O_n\cap\phi(\mathcal{M}))$.  Clearly these sets are open in the strongly attached point topology on $\overline{\mathcal{M}}$ as they are elements of the collection of basis sets $\mathcal{W}$. Every open set $V$ in $\overline{\mathcal{M}}$ that contains $[p]$ is an arbitrary union of $A_i=U_i\cup B_i$ sets.  One of the $B_j$ sets therefore contains $[p]$, and so $[p]$ is strongly attached to $U_j$. There thus exists an open neighbourhood $N$ of $p$ in $\widehat{\mathcal{M}}$ such that $N\cap\phi(\mathcal{M})\subseteq\phi(U_j)$. Now, there exists an $n\in\mathds{N}$, such that $\overline{O_n}\subset N$, and so $[\phi^{-1}(O_n\cap\phi(\mathcal{M}))]\cup B_{O_n}\subseteq V$.  This means that $\overline{\mathcal{M}}$ is first countable at $[p]$, for all $[p]\in\mathcal{B}(\mathcal{M})$.

We have thereby shown that the strongly attached point topology for $\overline{\mathcal{M}}$ is first countable.  $\Box$

\section{Optimal Embeddings and Partial Cross Sections}\label{optimal embeddings and partial cross sections}

When presented with a solution to the Einstein field equations in a particular coordinate system, it is not necessarily the case that these coordinates properly display all of its global and physical properties.  In practice, this often amounts to determining if the space-time is a proper subset of another, larger space-time.  The abstract boundary is therefore the natural boundary construction to use when considering extensions to space-times, given its utility in dealing with multiple envelopments at once.  An envelopment in which all of the global features of a space-time are evident may therefore be referred to as an optimal embedding.

In order to be able to choose an envelopment in which all of the global features of a space-time are properly displayed, the structure of the abstract boundary must be understood.  If a boundary set of an abstract boundary point is present in an envelopment, then we would like to know how the abstract boundary point represented by this boundary set is related to other abstract boundary points.  More specifically, we seek to know things like: is the abstract boundary point represented by that boundary set contained in some other abstract boundary point in some sense, i.e., is the abstract boundary point covered by some other abstract boundary point?  And therefore, is there a better, more complete way of displaying the boundary of the space-time in an envelopment? If there exists an envelopment in which more topological and physical information can be displayed, then clearly we should choose that envelopment.  Understanding the contact properties between abstract boundary points is therefore essential when considering optimal embeddings.

The contact properties that were defined earlier (definition \ref{abstract contact} and definition \ref{abstract seperate}) may be used to define subsets of $\mathcal{B}(\mathcal{M})$ referred to as partial cross sections.  Partial cross sections provide us with a way of abstracting the idea of envelopments as pictures of the boundary.  The abstract boundary is a very large object.  In some sense, the complete abstract boundary of a manifold $\mathcal{M}$ contains too much information. As discussed previously, different abstract boundary points can share large amounts of the same topological information.  It is therefore not necessary to consider every abstract boundary point in order to understand the structure of the abstract boundary. A partial cross section is a `slice' through the abstract boundary containing only abstract boundary points which are topologically distinct from each other. Partial cross sections are therefore important because, ideally, they can be used to simplify the abstract boundary to something more manageable.  In turn this can lead to the realisation of optimal embeddings. For further details on optimal embeddings see \cite{Ashley}.

\begin{definition}[Partial cross section $\sigma$] Let $\sigma\subset\mathcal{B}(\mathcal{M})$. $\sigma$ is a \textit{partial cross section} if for every $[p]$, $[q]\in\sigma$, $[p]\parallel[q]$ or $[p]=[q]$.\end{definition}

Of particular interest are partial cross sections of the following form:

\begin{example}Each envelopment $(\mathcal{M},\widehat{\mathcal{M}},\phi)$ defines a partial cross section $$\sigma_\phi:=\{[p]\mid p\in\partial(\phi(\mathcal{M}))\}.$$\end{example}

These $\sigma_\phi$ sets are important because we know what the topology on these sets should look like. Each abstract boundary point in $\sigma_\phi$ has a boundary point representative in the topological boundary $\partial(\phi(\mathcal{M}))$. The topology of this set is well defined by the topology on $\widehat{\mathcal{M}}$ and agrees with the relative topology on $\overline{\phi(\mathcal{M})}$, and hence it also agrees with the topology on $\mathcal{M}$ by virtue of the embedding $\phi$. Each $\sigma_\phi$ therefore has a natural topology defined on it by the given envelopment $(\mathcal{M},\widehat{\mathcal{M}},\phi)$.

\begin{definition}\label{partial topology}Let $\phi:\mathcal{M}\rightarrow\widehat{\mathcal{M}}$ be an envelopment, and $\sigma_{\phi}$ the partial cross section induced by $\phi$.  A natural topology $\mathcal{T}_{\sigma_\phi}$ is defined upon $\sigma_\phi$ by the topology of $\widehat{\mathcal{M}}$.  Let $N$ be an open neighbourhood of $\widehat{\mathcal{M}}$.  We then take a set $U$ to be an open set of $\sigma_{\phi}$ ($U\in\mathcal{T}_{\sigma_\phi}$) if and only if $U=\{[p]\in\sigma_\phi:\textrm{the singleton representative boundary point }p\in\partial(\phi(\mathcal{M}))\textrm{ is an element of }N\cap\partial(\phi(\mathcal{M}))\}$ for some open neighbourhood $N$.\end{definition}

As mentioned previously, the topology on $\partial(\phi(\mathcal{M}))$ is that induced by the topology on $\widehat{\mathcal{M}}$. Because the elements of $\sigma_\phi$ and $\partial(\phi(\mathcal{M}))$ are in one-to-one correspondence with each other, it follows that the collection $\mathcal{T}_{\sigma_\phi}$ of open sets of $\sigma_\phi$ given by definition \ref{partial topology} is indeed a topology on $\sigma_\phi$.

\begin{lemma}Let $\phi:\mathcal{M}\rightarrow\widehat{\mathcal{M}}$ be an envelopment, and $\sigma_\phi$ the partial cross section induced by $\phi$. The topological space $(\sigma_\phi,\mathcal{T}_{\sigma_\phi})$ is Hausdorff.\end{lemma}

\textit{Proof:} Let $[p]$, $[q]\in\sigma_\phi$, $[p]\neq[q]$, where $p$ and $q$ are distinct boundary points of $\partial(\phi(\mathcal{M}))$. Since the topology of $\widehat{\mathcal{M}}$ is Hausdorff, there exist disjoint open neighbourhoods $U$ and $V$ of $p$ and $q$, respectively, in $\widehat{\mathcal{M}}$. Now if we define $U^*=\phi^{-1}(U\cap\phi(\mathcal{M}))$ and $V^*=\phi^{-1}(V\cap\phi(\mathcal{M}))$, it follows that $U^*\cap V^*=\emptyset$, and $[p]$ is strongly attached to $U^*$ and $[q]$ is strongly attached to $V^*$. Define $A_{U^*}=U^*\cup B_{U^*}$ and $A_{V^*}=V^*\cup B_{V^*}$, where $B_{U^*}$ is the set of all abstract boundary points in $\sigma_\phi$ which are strongly attached to $U^*$ (so $[p]\in B_{U^*}$) and $B_{V^*}$ is the set of all abstract boundary points in $\sigma_\phi$ which are strongly attached to $V^*$ (so $[q]\in B_{V^*}$). Thus $A_{U^*}\cap\sigma_\phi=B_{U^*}$ and $A_{V^*}\cap\sigma_\phi=B_{V^*}$ are open sets of $\mathcal{T}_{\sigma_\phi}$ and open neighbourhoods of $[p]$ and $[q]$ respectively. Consider some $[r]\in\sigma_\phi$, where $r\in\partial(\phi(\mathcal{M}))$ and $r\neq p$, such that $[r]\in B_{V^*}$. There therefore exists an open neighbourhood $W$ of $r$ in $\widehat{\mathcal{M}}$ such that $W\cap\phi(\mathcal{M})\subseteq \phi(V^*)$. Now assume that $[r]\in B_{U^*}$. This implies that there exists an open neighbourhood $X$ of $r$ in $\widehat{\mathcal{M}}$ such that $X\cap\phi(\mathcal{M})\subseteq \phi(U^*)$. Since $U^*\cap V^*=\emptyset$ and $W\cap\phi(\mathcal{M})\subseteq \phi(V^*)$ and $X\cap\phi(\mathcal{M})\subseteq \phi(U^*)$, it follows that $[X\cap\phi(\mathcal{M})]\cap[W\cap\phi(\mathcal{M})]=\emptyset$. We also have that $X$ and $W$ are both open neighbourhoods of $r$, and so $[X\cap\phi(\mathcal{M})]\cap[W\cap\phi(\mathcal{M})]\neq\emptyset$. We therefore have a contradiction. This implies that $B_{U^*}$ and $B_{V^*}$ are disjoint open neighbourhoods of $[p]$ and $[q]$ respectively, thereby demonstrating that the topological space $(\sigma_\phi,\mathcal{T}_{\sigma_\phi})$ is Hausdorff. $\Box$
\\

In practice, the abstract boundary of a space-time is studied by considering its envelopments.  It is therefore highly desirable that the natural topology $\mathcal{T}_{\sigma_\phi}$ of a partial cross section $\sigma_\phi$ agrees with the topology on $\sigma_\phi$ induced by the strongly attached point topology.  That way, the topological features of the boundary may be studied in the natural topology of an envelopment and any results shown to be true in that envelopment will also hold in the strongly attached point topology on the whole abstract boundary $\mathcal{B}(\mathcal{M})$.

In the following proposition we show, assuming a condition holds, that the natural topology $\mathcal{T}_{\sigma_\phi}$ of a partial cross section $\sigma_\phi$ agrees with the topology on $\sigma_\phi$ induced by the strongly attached point topology.

\begin{figure}[htb!]
\centering%
\psfrag{manifold}{$\phi(\mathcal{M})$}
\psfrag{b}{$V$}
\psfrag{c}{$\textrm{congruence of curves }\{\lambda_p\}$}
\psfrag{a}{$\widehat{\mathcal{M}}$}
\psfrag{d}{$\partial(\phi(\mathcal{M}))$}
\includegraphics{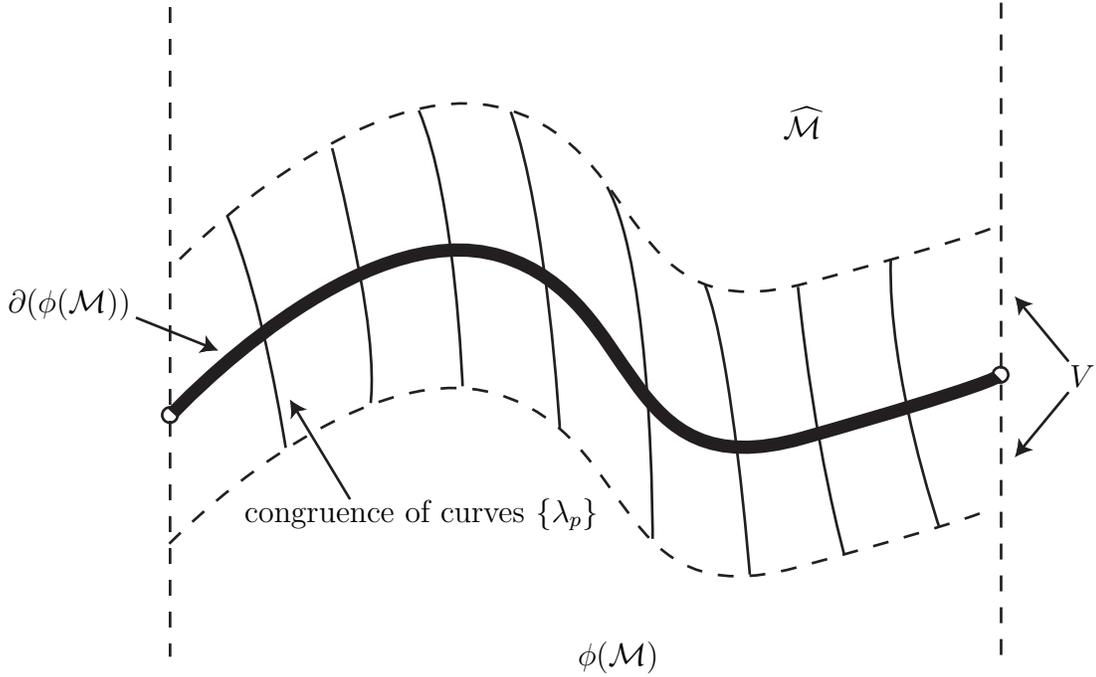}
\caption{an example of an envelopment $(\mathcal{M},\widehat{\mathcal{M}},\phi)$ which satisfies condition \ref{curve condition}.}
\label{condition iv}
\end{figure}

\begin{condition}\label{curve condition}Consider an envelopment $(\mathcal{M},\widehat{\mathcal{M}},\phi)$ with boundary $\partial(\phi(\mathcal{M}))\neq\emptyset$. There exists an open neighbourhood $V$ of $\overline{\partial(\phi(\mathcal{M}))}$ in $\widehat{\mathcal{M}}$ and a $C^2$ congruence of curves $\{\lambda_p\}$ on $V$ such that: \begin{enumerate} \item $\lambda_p$ passes through $p\in\overline{\partial(\phi(\mathcal{M}))}$ from one side to the other side of $\overline{\partial(\phi(\mathcal{M}))}$ where it exists as a surface

\item $\lambda_p\cap\overline{\partial(\phi(\mathcal{M}))}=\{p\}$

\item $\{\lambda_p\}$ is non-intersecting

\item For each $p$, $\lambda_p:(-\alpha,\beta)\rightarrow V$, where $\alpha,\beta\in\mathbb{R}^+$, such that: $\lambda_p(0)=p$, $\lambda_p(-\alpha)$, $\lambda_p(\beta)\in\overline{V}\backslash V$, $\lambda_p(-\alpha,0)\subset\phi(\mathcal{M})$, $\lambda_p(0,\beta)\subset\widehat{\mathcal{M}}\backslash\overline{\phi(\mathcal{M})}$ or $\lambda_p(0,\beta)\subset\phi(\mathcal{M})$.                                                 \end{enumerate}

See figure \ref{condition iv}. The possibility that $\lambda_p(0,\beta)\subset\phi(\mathcal{M})$ is included in (iv) to cover the case where, for a sufficiently small open neighbourhood $U$ of $p$ in $\widehat{\mathcal{M}}$, $U\backslash\partial(\phi(\mathcal{M}))\subseteq\phi(\mathcal{M})$.
\end{condition}

\begin{proposition}Let $T_{\sigma_\phi}$ be the topology on $\sigma_\phi$, defined by the topology of $\widehat{\mathcal{M}}$, given in definition \ref{partial topology}, and let $T_{\sigma_\phi(str)}$ be the topology on $\sigma_\phi$ induced by the strongly attached point topology on $\overline{\mathcal{M}}$. If $(\mathcal{M},\widehat{\mathcal{M}},\phi)$ obeys condition \ref{curve condition}, then $T_{\sigma_\phi}=T_{\sigma_\phi(str)}$.\end{proposition}

\textit{Proof:} $1)$ If $U$ is an open set of $T_{\sigma_\phi}$, then $U$ is an open set of $T_{\sigma_\phi(str)}$.
\\

Let $U$ be an open set of $T_{\sigma_\phi}$. This means that for some open set $N$ of $\widehat{\mathcal{M}}$ such that $N\cap\partial(\phi(\mathcal{M}))\neq\emptyset$, $U=\{[p]\in\sigma_\phi:p\in N\cap\partial(\phi(\mathcal{M}))\}$. It is clear that for each $p\in N\cap\partial(\phi(\mathcal{M}))$, $[p]$ is strongly attached to $\phi^{-1}(N\cap\phi(\mathcal{M}))$

We now consider whether any other abstract boundary points in $\sigma_\phi$ are strongly attached to $\phi^{-1}(N\cap\phi(\mathcal{M}))$. Suppose $q\in\partial(\phi(\mathcal{M}))\backslash\overline{N\cap\partial(\phi(\mathcal{M}))}$. Thus $q\in\widehat{\mathcal{M}}\backslash\overline{N}$ which is an open set disjoint from the open set $N$. Every open neighbourhood of $q$ will have non-empty intersection with $\widehat{\mathcal{M}}\backslash\overline{N}\cap\phi(\mathcal{M})$ and so $[q]$ is not strongly attached to $\phi^{-1}(N\cap\phi(\mathcal{M}))$.

Now suppose that $q\in\overline{N\cap\partial(\phi(\mathcal{M}))}\backslash N\cap\partial(\phi(\mathcal{M}))$. The envelopment $(\mathcal{M},\widehat{\mathcal{M}},\phi)$ obeys condition \ref{curve condition}, and so there exists an open neighbourhood $V$ of $\overline{\partial(\phi(\mathcal{M}))}$ in $\widehat{\mathcal{M}}$ which satisfies condition \ref{curve condition}. Consider the set $Y=\overline{\bigcup_p\lambda_p}$ where $p\in\overline{N\cap\partial(\phi(\mathcal{M}))}\backslash N\cap\partial(\phi(\mathcal{M}))$. Now define the set $N^*=\widehat{\mathcal{M}}\backslash Y\cap N\cap\phi(\mathcal{M})$. Clearly $N^*$ is an open subset of $\phi(\mathcal{M})$. Consider the point $q\in N\cap\partial(\phi(\mathcal{M}))$. Since $N$ is an open set, there exists a small open neighbourhood $N_q$ of $q$ in $N$ such that $N_q\cap\partial(\phi(\mathcal{M}))\subseteq N$. By condition \ref{curve condition}, a small open ball $B_\epsilon$ of radius $\epsilon$ about $q$ can be chosen such that $B_\epsilon\subseteq N_q$ and no curve $\lambda_p$ enters $B_\epsilon$, where $p\in\overline{N\cap\partial(\phi(\mathcal{M}))}\backslash N\cap\partial(\phi(\mathcal{M}))$. Now $B_\epsilon\cap\phi(\mathcal{M})\subseteq N^*$ and thus $N^*$ is non-empty. It follows that $[q]$ is strongly attached to $\phi^{-1}(N^*)$ for all $q\in N\cap\partial(\phi(\mathcal{M}))$. If $q\in\overline{N\cap\partial(\phi(\mathcal{M}))}\backslash N\cap\partial(\phi(\mathcal{M}))$, the curve $\lambda_q$ enters every open neighbourhood of $q$, and so $[q]$ is not strongly attached to $\phi^{-1}(N^*)$. If $q\in\partial(\phi(\mathcal{M}))\backslash\overline{N\cap\partial(\phi(\mathcal{M}))}$ we know that $[q]$ is not strongly attached to $\phi^{-1}(N\cap\phi(\mathcal{M}))$ and since $N^*\subseteq N\cap\phi(\mathcal{M})$, $[q]$ is not strongly attached to $\phi^{-1}(N^*)$. Thus $U$ is the open set of all abstract boundary points in $\sigma_\phi$ which are strongly attached to the open set $\phi^{-1}(N^*)$ of $\mathcal{M}$ and so $U$ is an open set of $T_{\sigma_\phi(str)}$.
\\

$2)$ If $U$ is an open set of $T_{\sigma_\phi(str)}$, then $U$ is an open set of $T_{\sigma_\phi}$.
\\

Let $U$ be an open set of $T_{\sigma_\phi(str)}$. There therefore exists an open set $\bigcup_i A_i=\bigcup_i U_i\cup B_i$ of $\overline{\mathcal{M}}$, where each $U_i$ is a non-empty open set of $\mathcal{M}$ and $B_i$ is the set of all abstract boundary points which are strongly attached to $U_i$, such that $U=(\bigcup_i U_i\cup B_i)\cap\sigma_\phi=(\bigcup_i B_i)\cap\sigma_\phi$.

Consider $[p]\in(\bigcup_i B_i)\cap\sigma_\phi$, where $p\in\partial(\phi(\mathcal{M}))$. There exists a $B_i$ such that $[p]\in B_i$ and thus $[p]$ is strongly attached to $U_i$. This means that there exists an open neighbourhood $N_p$ of $p$ in $\widehat{\mathcal{M}}$ such that $N_p\cap\phi(\mathcal{M})\subseteq\phi(U_i)$. Consider a boundary point $q\in\partial(\phi(\mathcal{M}))$ such that $q\in N_p$. It is clear that $[q]\in B_i$ and thus $[q]\in U$. The set $W=\bigcup_{[p]\in U} N_p$ is an open set in $\widehat{\mathcal{M}}$. Consider the open set in $T_{\sigma_\phi}$ defined by $A=\{[p]:p\in W\cap\partial(\phi(\mathcal{M}))\}$. It is clear that $U\subseteq A$ and, from the above argument, that $A\subseteq U$. Thus $U$ is an open set of $T_{\sigma_\phi}$. $\Box$
\\

Direction $2$ of the previous proof is quite straightforward. Direction $1$ on the other hand, is more complicated and requires us to invoke condition \ref{curve condition}. We have to use this condition due to the existence of boundary points $p\in\partial(\phi(\mathcal{M}))$ that are strongly attached to $\phi^{-1}(N\cap\phi(\mathcal{M}))$ but are not elements of $N$. The existence of these boundary points makes it difficult to construct an open neighbourhood of $T_{\sigma_\phi(str)}$ that doesn't contain abstract boundary points additional to those contained in $U$. Even so, condition \ref{curve condition} is not very restrictive and may even hold in general. At the least, we have been unable to construct a space-time in which it does not hold.

\section{Conclusion}

There are many topologies that can be placed on $\overline{\mathcal{M}}$. They will not all be physically useful, however. Ultimately, a topology should provide a structure for $\overline{\mathcal{M}}$ which aids us in answering physical questions about $\overline{\mathcal{M}}$.  Ideally then, the topology should connect the abstract boundary to the manifold in a physically meaningful way, and the resulting structure on $\overline{\mathcal{M}}$ should conform to many of our intuitive ideas regarding the behaviour of `missing points', i.e., abstract boundary points, from the manifold $\mathcal{M}$.

The strongly attached point topology was defined similarly to the attached point topology but with one important difference. This difference, related to the way in which abstract boundary points are `attached' to open sets of $\mathcal{M}$, means that the strongly attached point topology does not need to include collections of abstract boundary points.  In the attached point topology these sets were necessary to ensure that the basis for the topology was well-defined.  In some sense, because the abstract boundary points are more firmly connected to the manifold in the strongly attached point topology, we avoid having to add more open sets to the topology.  It is interesting to note that as a consequence of this, every open neighbourhood of an abstract boundary point in the strongly attached point topology necessarily contains some part of the manifold $\mathcal{M}$, thereby encapsulating the true essence of a boundary point.

Another consequence of the strongly attached point topology not containing collections of abstract boundary points is that there exist abstract boundary points which are not Hausdorff separated from each other.  While it has been argued that a physically useful topology for a space-time should be Hausdorff \cite{hajicek}, the lack of Hausdorff separation between abstract boundary points in the strongly attached point topology, nevertheless, contains useful information about the boundary. It was demonstrated that two abstract boundary points are Hausdorff separable if and only if they are not in contact. Intuitively, this makes sense as two abstract boundary points which are in contact share much of the same topological information, and therefore they do not represent two points which are distinct from each other. It therefore seems reasonable that abstract boundary points which are in contact with each other cannot be separated by disjoint open sets. It is also worth noting that there is a natural relationship between the separation axioms that two abstract boundary points obey, and how much topological information they share. As propositions \ref{hausdorff separated} through \ref{T0 sep1} show, as two abstract boundary points share more of the same topological information, they obey fewer separation axioms. For example, two abstract boundary points which are in contact are $T_1$ separable, but not $T_2$ separable,  while if one abstract boundary point covers the other, they are $T_0$ separable, but not $T_1$ separable. Therefore, while separation is lost between abstract boundary points, it is lost in a way directly related to the amount of overlap between the abstract boundary points.

The strongly attached point topology possesses a number of other interesting properties which suggest that it is an appropriate topology for $\overline{\mathcal{M}}$.  One such property is that the description of $\mathcal{M}$ and $\mathcal{B}(\mathcal{M})$ in the strongly attached point topology agrees with many of our intuitive ideas about the nature of a space and its boundary.  Traditionally, singularities are typically viewed as `points' missing from a space-time.  We can approach these missing points from within the space-time, becoming arbitrarily close to them, but we cannot reach them.  In some sense then, these missing points make up the `closure' of the space-time, and ideally, a topology on $\overline{\mathcal{M}}$ should reflect this.  The strongly attached point topology agrees with this notion in the sense that $\mathcal{M}$ is open and not closed, $\mathcal{B}(\mathcal{M})$ is closed and not open, and $\mathcal{M}$ can be embedded identically into $\overline{\mathcal{M}}$.  The strongly attached point topology therefore provides a natural way of viewing the structure of $\overline{\mathcal{M}}$ in that it can be seen as $\mathcal{M}$ with the addition of a topological boundary made up of abstract boundary points.

Perhaps the most important property of the strongly attached point topology is that the topology induced by the strongly attached point topology on a partial cross section $\sigma_\phi$ associated with an envelopment $\phi:\mathcal{M}\rightarrow\widehat{\mathcal{M}}$ generally agrees with the natural topology on $\sigma_\phi$.  In practice, the abstract boundary is studied via envelopments of the manifold $\mathcal{M}$.  Consequently, the embedded manifold $\phi(\mathcal{M})$ and its topological boundary $\partial(\phi(\mathcal{M}))$ already have a topology defined on them with which it is very easy to work.  It is therefore very useful that the topologies agree on the partial cross sections $\sigma_\phi$ as it means that any topological result which holds in an envelopment (which, again, is where the abstract boundary is studied in practice) will also hold with respect to the larger topology on $\mathcal{B}(\mathcal{M})$.

Without a topology on $\overline{\mathcal{M}}$ we cannot say `where' singular points are with respect to the manifold $\mathcal{M}$. A topology on $\overline{\mathcal{M}}$ should therefore relate the abstract boundary back to the manifold $\mathcal{M}$.  Moreover, it should ideally do so in a natural way, i.e., the topology should describe the singular points in a way that agrees with our intuitive ideas of how a singularity is related to the manifold.  It has been shown that the strongly attached point topology does indeed relate the abstract boundary back to the manifold in a way that encompasses many of our intuitive notions of the nature of a topological boundary. For these reasons, the strongly attached point topology appears to be a particularly good choice for a topology on the set comprising a manifold and its abstract boundary.
\\

\textbf{Acknowledgements}
\\

We would like to thank the referees for their detailed and very helpful comments. We would also like to thank Ben Whale for his suggestions and the discussion of numerous ideas.

\end{document}